\documentclass[onecolumn]{emulateapj}

\def\etal{{et~al. \,}}
\def\mo{M$_\odot$}

\def\kms{\,km\,s$^{-1}$}
\def\mdot{\,M$_\odot$\,s$^{-1}$\,}
\def\undertext#1{$\underline{\smash{\hbox{#1}}}$}
\def\sles{\lower2pt\hbox{$\buildrel {\scriptstyle <}
   \over {\scriptstyle\sim}$}}

\def\sgreat{\lower2pt\hbox{$\buildrel {\scriptstyle >}
   \over {\scriptstyle\sim}$}}
\def\sharpnull#1{}
\def\aa{Astron. Astrophys.\ }

\def \BF  {\mathbf F}
\def \Bg  {\mathbf g}
\def \half  {{1 \over 2}}

\def \Gpi {{1 \over 4\pi G }}
\def \dJgdt {{1 \over c} {\partial J_g \over {\partial t}}}
\def \cdt   {{1 \over {c \Delta t}}}
\def \dsi  {\mathbf {dS_i}}
\def \vel {{\mathbf v}}
\def \dt    {\Delta t}
\def \pos {{\mathbf r}}
\def \ovr  {\frac{1}{r}}

\begin{document}

\slugcomment{\bf}
\slugcomment{Accepted to Ap.J. September 29, 2006}

\title{Features of the Acoustic Mechanism of Core-Collapse Supernova Explosions}

\author{A. Burrows\altaffilmark{1},  
E. Livne\altaffilmark{2},
L. Dessart\altaffilmark{1},
C.D. Ott\altaffilmark{3},
J. Murphy\altaffilmark{1}}
\altaffiltext{1}{Department of Astronomy and Steward Observatory, 
                 The University of Arizona, Tucson, AZ \ 85721;
                 burrows@as.arizona.edu,luc@as.arizona.edu,jmurphy@as.arizona.edu}
\altaffiltext{2}{Racah Institute of Physics, The Hebrew University,
Jerusalem, Israel; eli@frodo.fiz.huji.ac.il}
\altaffiltext{3}{Max-Planck-Institut f\"{u}r Gravitationsphysik,
Albert-Einstein-Institut, Golm/Potsdam, Germany; cott@aei.mpg.de}

\begin{abstract}

In the context of 2D, axisymmetric, multi-group, radiation/hydrodynamic 
simulations of core-collapse supernovae over the full 180$^{\circ}$ domain, 
we present an exploration of the progenitor dependence of the acoustic 
mechanism of explosion.  All progenitor models we have tested with our 
Newtonian code explode. However, some of the cores left behind in our simulations,
particularly for the more massive progenitors, have baryon masses that 
are larger than the canonical $\sim$1.5 M$_{\odot}$ of well-measured 
pulsars. We investigate the roles of the Standing-Accretion-Shock-Instability 
(SASI), the excitation of core g-modes, the generation of core acoustic power, 
the ejection of matter with r-process potential, the wind-like character of 
the explosion, and the fundamental anisotropy of the blasts.  We find that the 
breaking of spherical symmetry is central to the supernova phenomenon, the 
delays to explosion can be long, the ejecta are radiation-dominated, and the 
blasts, when top-bottom asymmetric, are self-collimating.  We see indications
that the initial explosion energies are larger for the more massive progenitors,
and smaller for the less massive progenitors, and that the neutrino contribution to 
the explosion energy may be an increasing function of progenitor mass.  
However, the explosion energy is still accumulating by the end of our
simulations and has not converged to final values. The degree of explosion 
asymmetry we obtain is completely consistent with that inferred from the 
polarization measurements of Type Ic supernovae. Furthermore, 
we calculate for the first time the magnitude and sign of the net impulse 
on the core due to anisotropic neutrino emission and suggest that hydrodynamic 
and neutrino recoils in the context of our asymmetric explosions afford a natural 
mechanism for observed pulsar proper motions. We conclude that mechanical 
and fundamentally hydrodynamic mechanisms of supernova explosion may provide 
viable alternatives to the canonical neutrino mechanism. We discuss the 
numerical challenges faced when liberating the core to execute its 
natural multi-dimensional motions in light of the constraints of momentum 
and energy conservation, the need to treat self-gravity conservatively, 
and the difficulties of multi-dimensional neutrino transfer.

\end{abstract}


\section{Introduction}
\label{intro}

One of the outstanding problems in astrophysics is the
mechanism of core-collapse supernova explosions.  This pedigreed
puzzle has resisted more than forty years of theoretical speculation
and numerical exploration.  It has intrigued many unaware of the 
numerous feedbacks that mute the consequences of most alterations 
in microphysical processes.  It has tricked specialists with a legendary
list of false starts and blind alleys.  It has taunted  
computational astrophysicists with both its imagined and real 
complexities.  The potential roles as factors in explosion 
of neutrinos, the nuclear equation of state (EOS), exotic physics,
general relativity, dimensionality, instabilities, magnetic
fields, and rotation all continue to be topical.

The neutrino-heating mechanism, in which a stalled bounce shock is reenergized
by neutrino energy deposition after a slight delay, perhaps aided by overturning
instabilities in this ``gain region," has been the working hypothesis of the community for
the last twenty years (Bethe \& Wilson 1985; Buras et al. 2006ab; 
Liebend\"{o}rfer et al. 2001).  Past calculations in support of
this mechanism, or variations on its theme, include those by Wilson \& Mayle (1988,1993),
Mayle \& Wilson (1988), Herant et al. (1994), Burrows, Hayes, \& Fryxell (1995), 
Janka \& M\"uller (1996), and Fryer \& Warren (2002,2004). 

Nevertheless, recent calculations employing careful
neutrino physics and numerics suggest that the neutrino mechanism, 
when it succeeds, may at best be marginal.
Kitaura et al. (2006) follow in spherical symmetry the 
compact 1.38-M$_{\odot}$ O-Ne-Mg core of the 8.8-M$_{\odot}$ model
of Nomoto \& Hashimoto (1988), with a very tenuous outer envelope, and 
obtain a delayed neutrino-driven explosion.
However, the explosion energy is only $\sgreat$10$^{50}$ ergs and the major driver of 
the explosion is the neutrino-driven wind (Burrows 1987; Burrows 
\& Goshy 1993; Janka et al. 2005ab).  Buras et al. (2006b)
witness the onset of the SASI-aided\footnote{SASI: \undertext{S}tanding 
\undertext{A}ccretion \undertext{S}hock \undertext{I}nstability; see Blondin, Mezzacappa, \& DeMarino 2003,
Blondin \& Mezzacappa 2006, Foglizzo \& Tagger 2000, Foglizzo 2001,2002,
Foglizzo, Galletti, \& Ruffert 2005, Buras et al. 2006ab, 
Burrows et al. 2006, and Foglizzo et al. 2006.}, 
neutrino-driven $\ell = 1$ explosion of the 
11.2-M$_{\odot}$ progenitor of Woosley, Heger, \& Weaver (2002, WHW02).
Curiously, they also infer a very weak explosion energy, this time near 10$^{49}$ ergs, not correcting for
neutrino driving subsequent to the early termination of their calculation or for the 
binding energy of the outer envelope.  Buras et al. (2006b) focus on the importance
of calculating over the full 180$^{\circ}$ 2D domain so as not to suppress the $\ell = 1$ mode
of the SASI, since they do not obtain even a weak explosion when 
constraining the computational domain to 90$^{\circ}$. They also imply 
that performing the calculations in 3D might make the explosion robust.  

However, perhaps the most interesting conclusion of the Buras et al. (2006b) paper is
that of all the progenitors they study (including 11.2-, 13-, 15-, 20-, and 25-M$_{\odot}$ models),
only the 11.2-M$_{\odot}$ star, with its steep outer 
density gradient, small iron core ($\sim$1.26 M$_{\odot}$),
and close-in Si/O density shelf, approaches the $\tau_{adv}\, \sim\, \tau_{H}$
condition, promoted by Thompson, Quataert, \& Burrows (2005)
as the litmus test for the neutrino-driven explosion mechanism\footnote{$\tau_{adv}$
is the timescale for shocked matter to traverse the gain region and $\tau_{H}$ is the neutrino heating timescale.
When $\tau_{adv}$ is long and/or $\tau_{H}$ is short, neutrino heating can explode the shocked mantle.}.
Furthermore, Buras et al. (2006b) find that only the 11.2-M$_{\odot}$ progenitor amply satisfies the mantle
overturn and perturbation growth condition (Foglizzo, Scheck, \& Janka 2006) they evoke to analyze the potential
for the SASI in the first $\sim$200 milliseconds after bounce. 

From the work of Kitaura et al. (2006), Buras et al. 
(2006b), and Janka et al. (2005ab) one might conclude that
the neutrino mechanism requires compact cores and tenuous outer envelopes,
but that for such progenitors a solely neutrino-driven explosion 
is perforce weak. The small accretion rates that may be
necessary after bounce for the neutrino mechanism to succeed
ensure that when it does succeed there is little mass to absorb
the driving neutrino luminosity, resulting in the weak explosion.
The neutrino-driven mechanism alone, and on its own, may not be able  
to yield a robust, $\sim$10$^{51}$-erg explosion.

The large mass accretion rates experienced by more massive cores after bounce
create more massive post-shock absorbing/gain regions and higher neutrino luminosities, but those
same accretion rates may well tamp and suppress the neutrino-driven explosion.
However, for the more massive progenitors
that don't now seem to explode by the neutrino mechanism, it may be that
better neutrino transfer, coupled with and aided by an $\ell = 1$ SASI mode calculated
over the full 180$^{\circ}$ domain, will follow the path of the 11.2-M$_{\odot}$
progenitor as simulated by Buras et al. (2006b) and result in explosions.
Three-dimensional simulations with sophisticated neutrino numerics and physics have
not been performed and may well reveal qualitative, or large quantitative, differences
with the results in two dimensions. The neutrino mechanism may yet prove to
be more muscular and universal. However, it is also possible that the typical
supernova of the typical massive-star progenitor does not explode by the neutrino
mechanism, or solely by the neutrino mechanism, and that another,
non-magnetic\footnote{Though we suspect that strong magnetic fields require
very rapid rotation that may not be available in the generic
core-collapse supernova context (Ott et al. 2006a), magnetic jets have been
suggested as potential power sources (Akiyama et al. 2003; Kotake et al. 
2006; Moiseenko et al. 2006; Obergaulinger et al. 2006).  Perhaps
rapid rotation could facilitate an MHD scenario for the rare hypernovae and
long/soft gamma-ray bursts (GRBs), or for accretion-induced collapse (Dessart et al. 2006a).},
mechanism may be at work.

Recently, Burrows et al. (2006) have proposed an acoustic mechanism
for exploding core-collapse supernovae.  In it, the progressive growth
of the SASI and of the entropy and Mach number of the accreted shocked
matter long after the outer shock has stalled results in anisotropic accretion onto the inner core
that over time excites core g-modes\footnote{The SASI itself is insufficient
to explode the supernova. Contrary to Blondin, Mezzacappa, \& DeMarino (2003),
when proper account is taken of the neutrino losses, the nuclear equation of state, and the
inner boundary, the total transverse turbulent kinetic energy in the shocked zone does not grow
(Burrows et al. 2006; Buras et al. 2006ab).}.  Predominantly $\ell = 1$
in character, these core eigenmodes achieve large amplitudes and dampen by
the radiation of sound.  Multiple sound pulses emanate from the
core with periods of 2-4 milliseconds and steepen into shock waves.
The resulting acoustic power deposits energy and momentum aspherically
into the outer shocked mantle and explodes the supernova,
but on timescales of many hundreds of milliseconds to seconds
\footnote{If another mechanism (such as the neutrino mechanism) were to explode
the envelope significantly earlier, the inner core oscillations might not 
be excited to importance and the core acoustic model might be aborted.}.
The blast is fundamentally aspherical, favoring one side.
During the early explosion, the other side continues to experience accretion,
which maintains the core oscillation and the generation of sound
until the entire mantle has exploded. Thus, the breaking of spherical
symmetry and the excitation and maintenance of aspherical g-modes allow simultaneous
accretion and explosion. As long as it is needed to ensure
success, the core acts like a transducer for the conversion
of accretion gravitational power into outwardly-propagating acoustic
power. Curiously, the recoil due to the resulting anisotropic mass loss pushes
the accretion streams that are exciting the core oscillation into
a configuration that is even more favorable for the excitation of $\ell = 1$ core
oscillations (\S\ref{discussion}; Burrows et al. 2006).  In this way, the core g-modes
seem self-excited.  This is what we observe in our simulations,
but such an intriguing phenomenon remains to be verified.  
Perhaps, even if the initial explosion is neutrino-driven
and it too is very asymmetrical due to the $\ell = 1$ SASI mode, 
the recoil due to the anisotropic neutrino-driven mass loss
can, by the same mechanism, excite core g-mode oscillations. Hence, even a neutrino-triggered explosion might 
excite core oscillations, which would radiate acoustic power and boost the explosion energy.
Such a ``hybrid" mechanism for supernova explosions is a particularly
intriguing possibility, but has yet to be adequately explored.

One may further speculate that there is not one 
core-collapse supernova mechanism, but several.  The lowest-mass
massive star progenitors (Kitaura et al. 2006; Buras et al 2006b)
and accretion-induced collapse (Dessart et al. 2006a) might explode
early and subenergetically by the neutrino-driven wind mechanism,
while the generic progenitor might explode by the acoustic mechanism (aided 
by neutrino heating).  However, when better neutrino transport than 
we have employed is included, we may find that the neutrino/acoustic hybrid mechanism
obtains throughout most of the progenitor mass range, particularly if the explosion
commences after the vigorous phase of the $\ell = 1$ SASI begins, with the relative contributions
of neutrino heating and acoustic power varying as a function of progenitor mass.
Finally, the subset with rapidly rotating cores, or cores that collapse into black
holes before explosion, might rely on MHD processes
to explode the mantle, and these might be associated with hypernova
and/or GRBs.  All in all, the outcome will depend upon the progenitor's
inner density and rotation profiles.

Our calculations support the notion that all non-rotating progenitors
that do not explode by the early neutrino mechanism experience the SASI,
and later excite core pulsations that generate acoustic power that aids
or enables explosion\footnote{Unless, the core first collapses 
to a black hole.}.  In this paper, we present the results of preliminary
investigations into the progenitor dependence of the core oscillation/acoustic mechanism.
In \S\ref{prog}, we compare the density profiles of representative non-rotating progenitor models
of Woosley \& Weaver (1995, WW95), Woosley, Heger, \& Weaver (2002), and Nomoto \& Hashimoto (1988)
and discuss the resulting mass accretion and protoneutron star mass accumulation histories.    
These profiles determine the outcome of collapse.
In \S\ref{evolution}, we discuss the overall hydrodynamic behavior of some of these models, 
focusing on the evolution of the shock position, the dependence of the SASI
frequency upon progenitor profile, and the core pulsation energy.
In \S\ref{entropy}, we discuss the entropies achieved, the degree of radiation domination,
and the possible consequences for the r-process.  We then go on in \S\ref{aniso-wind} to 
examine the aspherical wind-like character of the explosion and its consequences and the anisotropy
of the neutrino emissions and the resulting recoils.  This leads us to suggest a natural model 
for pulsar kicks.  Finally, in \S\ref{discussion} we summarize and discuss 
our conclusions concerning the core-oscillation/acoustic 
mechanism of core-collapse supernova explosions.  

For these simulations, we have used the two-dimensional
multi-group, multi-neutrino-species, flux-limited diffusion (MGFLD) variant of the
code VULCAN/2D (Livne et al. 2004; Walder et al. 2005; Ott et al. 2006ab;
Dessart et al. 2006ab; Burrows et al. 2006) and describe many of its numerical features in the Appendix.
This is currently the only extant 2D, multi-group code that allows core translational 
motion by introducing a Cartesian-like grid in the inner core and, hence, that
is capable of investigating the core-oscillation/acoustic mechanism. VULCAN/2D
is also the only extant supernova code to perform 2D (not ``ray-by-ray") multi-group transport.  Due to the 
finite-difference character of 2D codes that employ spherical coordinates 
all the way to the center, to the singularity in those coordinates at that center,
and to the reflecting boundary condition frequently imposed at this center,
spherical-coordinate codes are likely to inhibit core translational motions 
artificially and, hence, to inhibit the $\ell = 1$ g-modes that are central to
the mechanism we have identified. Be that as it may, 
there are many caveats to our study the reader should keep in mind:  1) Our 
calculations are Newtonian and not general-relativistic, 2) As stated above, we 
employ an approximate multi-group transport algorithm in the neutrino sector, 3) Numerical errors
are bound to have accumulated due to the need to calculate for $\sim$1,000,000 timesteps 
for each progenitor, and 4) The initial seed perturbations are unknown (and unknowable?). 
Furthermore, the flow is fundamentally chaotic and a precise mapping between initial
configuration and final outcome is not possible.  This multi-dimensional radiation/hydrodynamical
problem is quintessentially meteorological in character.  Nevertheless, along with
the work of Burrows et al. (2006), these are the first calculations to explore
the novel core oscillation/acoustic mechanism and to venture into the 
late-time behavior of multi-D core collapse with multi-D core motions
and multi-D/multi-group transport.

\section{Progenitor Density Profiles and Mass Accumulation Rates}
\label{prog}

The basic evolutionary phases through which a core proceeds
in the context of the core-oscillation/acoustic supernova mechanism have been described in
Burrows et al. (2006), to which the reader is referred for details.  These are 
summarized in \S\ref{intro} and in \S\ref{discussion}. Burrows et al. (2006) explored
the results for the 11-M$_{\odot}$ model of Woosley \& Weaver (1995) alone.  Since that paper,
we have calculated more models, including the 25-M$_{\odot}$ model of Woosley \& Weaver (1995)
and the rotating m15b6 model of Heger et al. (2005) (Ott et al. 2006ab), as well 
as the 11.2-M$_{\odot}$, 13-M$_{\odot}$, 15-M$_{\odot}$, 20-M$_{\odot}$, and 
25-M$_{\odot}$ progenitor models of WHW02 and the 13-M$_{\odot}$ and 15-M$_{\odot}$
models of Nomoto \& Hashimoto (1988).  All models explode, modulo
any fallback at very, very late times not yet accessible to supernova codes.  This set of models
constitutes the most extensive and detailed radiation/hydrodynamic study of the shock instability (SASI) 
and of the multi-dimensional core motions undertaken to date.   

\begin{figure*} 
\figurenum{1}
\plotone{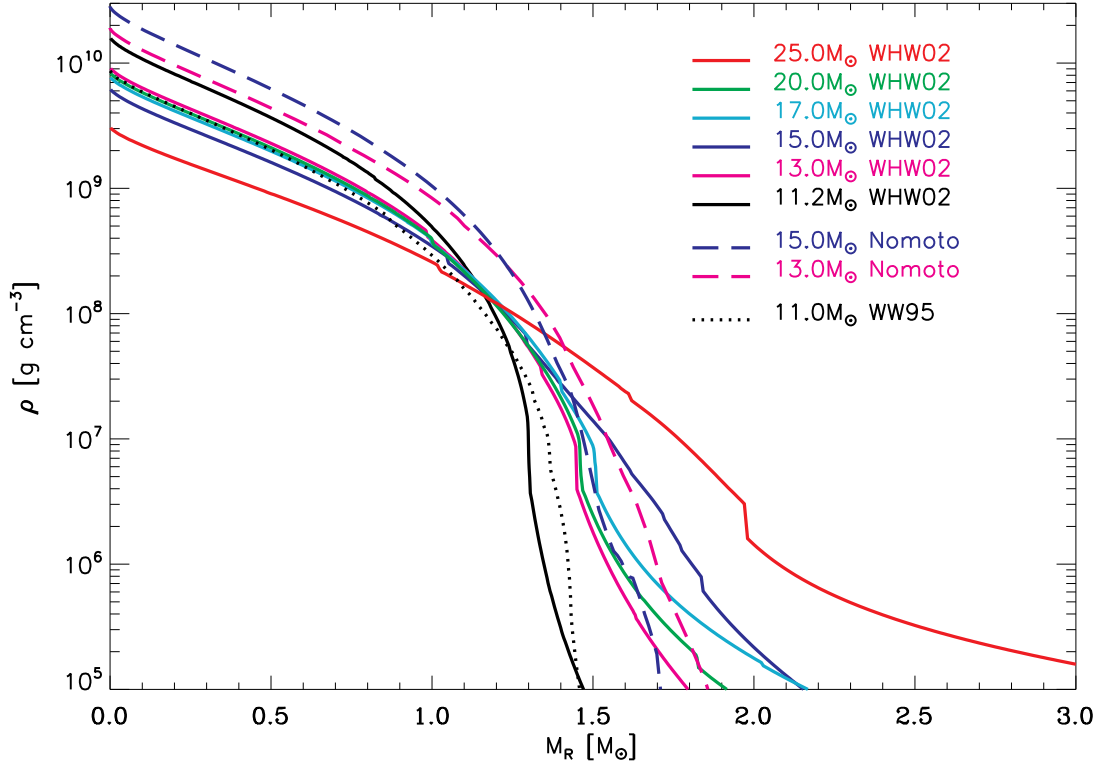}
\caption{
Profiles of the mass density (in units of g cm$^{-3}$) versus Lagrangean mass 
(in \mo) of representative massive star progenitor cores of Woosley, Heger, \& Weaver (2002 --
11.2\,\mo: black; 13\,\mo: magenta; 15\,\mo: blue;
17\,\mo: turquoise; 20\,\mo: green; 25\,\mo: red),
Nomoto \& Hashimoto (1988 -- 13\,\mo: magenta; 15\,\mo: blue), 
and Woosley \& Weaver (1995 -- 11\, \mo: dotted).  See text for
a discussion on the import of these profiles.
}
\label{fig:1}
\end{figure*}

For non-rotating models, the most important determinant of the outcome of collapse
is the density profile in the inner thousands of kilometers of the massive star 
progenitor.  The structure of this ``Chandrasekhar" core, with surrounding inner envelope,  
is determined by the burning history to the point of instability.  This history
reflects the various core and shell burning stages, and is a function in Nature
of ZAMS mass, mass loss, and metallicity.  However, different theoretical groups performing
calculations of the evolution of massive stars and using different
approaches to semi-convection, overshoot, convection, and mass loss still do not end 
with the same configurations.  Figure \ref{fig:1} provides some density profiles
for progenitor models from Woosley, Heger, \& Weaver (2002), Nomoto \& Hashimoto (1988),
and Woosley \& Weaver (1995) at a point just after collapse ensues.  The first
thing to note is that there is a spread in structures and that the ``Chandrasekhar"
core is not the same for all progenitors, but varies in structure and mass.  The corresponding 
Y$_e$ and entropy profiles vary similarly.  Secondly, as a comparison of the two sets of 13-M$_{\odot}$
and 15-M$_{\odot}$ models shown on Fig.~\ref{fig:1} makes clear, the structure for
a given progenitor mass has not converged theoretically.  Different groups arrive
at different profiles for the same ZAMS mass.  Thirdly, the density profiles are   
not necessarily monotonic with ZAMS progenitor mass: the 15-M$_{\odot}$ model of 
WHW02 has a shallower profile than that of their
20-M$_{\odot}$ model and the 15-M$_{\odot}$ model of Nomoto \& Hashimoto (1988)
has a steeper profile than that of their 13-M$_{\odot}$ model.  
Finally, the older 11-M$_{\odot}$ model of WW95 and the
more recent one of WHW02 at 11.2 M$_{\odot}$, while both being steep, are not equally
steep in the same regions.  The 11.2-M$_{\odot}$ model has lower densities between
interior masses from 1.2 M$_{\odot}$ to 1.45 M$_{\odot}$, while the 11-M$_{\odot}$
model of WW95 has lower densities exterior to that mass (not shown).  The upshot 
is that the outcome of collapse and the character of whatever explosion is ignited
are not likely to be the same.  In particular, the 11.2-M$_{\odot}$ model of WHW02
boasts the thinnest mantle of their whole model set, and this is consistent with the
explanation given in \S\ref{intro} for why Buras et al. (2006b)
obtained a SASI- and neutrino-aided explosion, albeit weak, but no such explosion for the more massive
cores with shallower and thicker density profiles.  The even-steeper profile of the ONeMg 
model of Nomoto \& Hashimoto (1988) (not shown on Fig. \ref{fig:1}) explains the
results of Kitaura et al. (2006) and the near-vacuum of the outer envelopes used
in the accretion-induced-collapse simulations of Dessart et al. (2006a) explains
why they saw weak neutrino-aided explosions.  

\begin{figure*} 
\figurenum{2}
\plotone{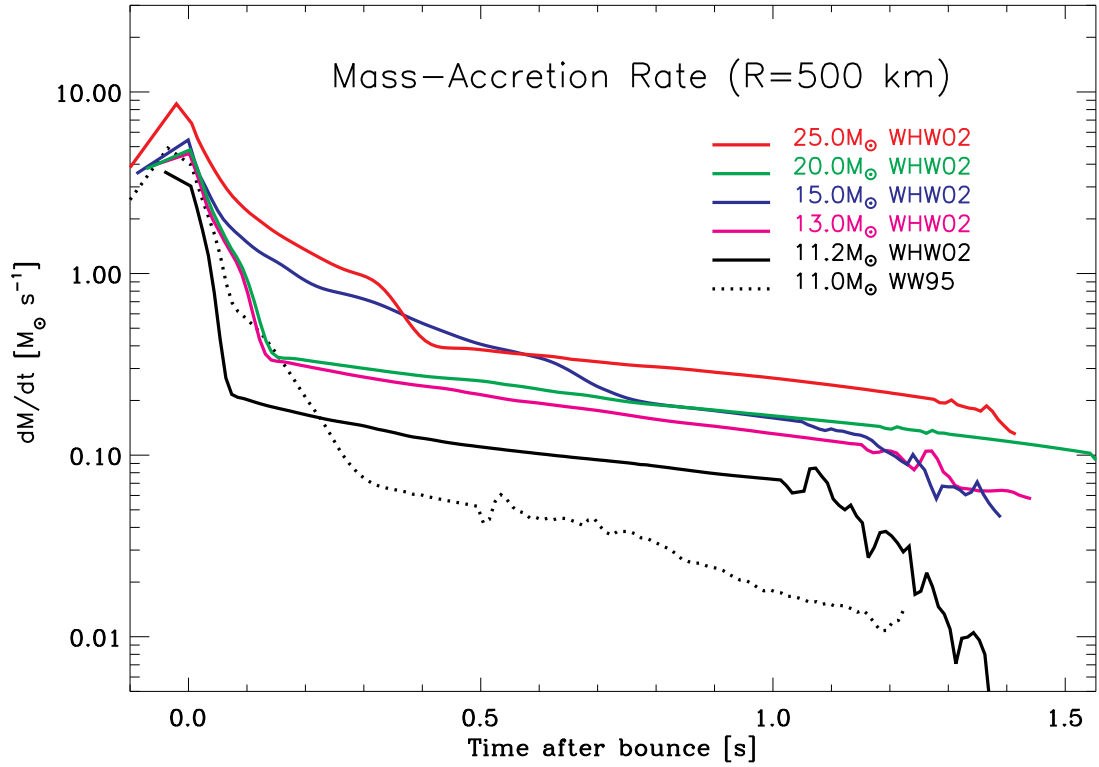}
\caption{
Pre- and post-bounce time evolution of the accretion rate through
a radius of 500 km (in \mo~s$^{-1}$), for various massive 
star progenitors.  Only the infalling matter is included.
The color coding and linestyles have been retained from 
Fig.~\ref{fig:1}.
}
\label{fig:2}
\end{figure*}

The structures depicted in Fig.~\ref{fig:1} translate directly into mass accretion
rates ($\dot{M}$s) through the stalled shock. Because the inner shocked region and the core
are out of sonic contact with this mantle, $\dot{M}$ and its evolution after bounce are functions of this
structure alone.  Hence, for diagnosing and ``predicting" the outcome of collapse,
the post-bounce behavior of $\dot{M}$ for a given progenitor is useful, and probably determinative.
Figure \ref{fig:2} portrays the evolution of the mass accretion rate for representative
progenitor models evolved using the 2D MGFLD variant of VULCAN/2D.  The wide range
of curves reflects the range of profiles plotted in Fig. \ref{fig:1}.  At 0.5 seconds after bounce,
$\dot{M}$ varies from 0.06 to 0.5 \mdot, while at 1.0 second it varies
from 0.02 to 0.3 \mdot, an order-of-magnitude span at both epochs.  A glance
at the behavior of $\dot{M}$ for the 11.2-M$_{\odot}$ model used in Buras et al. (2006b)
shows its steep drop at early times, and the corresponding lower accretion tamp. 
Such a precipitous drop is not in evidence for the other, more massive, 
WHW02 progenitors portrayed in Fig. \ref{fig:2}.

The plummeting of $\dot{M}$ at later times for the 11-M$_{\odot}$ model of WW95
used in Burrows et al. (2006) is tied to the earlier onset of the sound-powered
explosion they witness than we generally find in this paper, employing as we
do here a fuller range of progenitor models with shallower density profiles.   
The drops in $\dot{M}$ seen at the latest times shown on Fig. \ref{fig:2} are consequences
of the late-onset acoustic-driven explosions we find, with the 11.2-M$_{\odot}$ model
exploding earliest.  Probably because our MGFLD neutrino transfer can still
be improved, we do not reproduce the weak neutrino-driven explosion seen 
by Buras et al. (2006b) for the 11.2-M$_{\odot}$ model. However, determining the precise
reasons for the difference will entail a direct comparison of the details of both 
codes, something that will be subtle.  As we show in Dessart et al. (2006b),
the luminosities and matter profiles are generally similar for the two codes.
The Buras et al. calculations were done using a GR substitute; ours
were Newtonian.  The Buras et al. calculation does a remap between
their Eulerian (PPM) hydro code and comoving-frame (Lagrangian)
transport.  Ours does Lagrangian hydro, and then remaps to an Eulerian
grid. Buras et al. calculate spherical transport along radial rays, using
for each ray the same Eddington factor for an average sphere.  We do
flux-limited diffusion, but in full 2D. Buras et al. don't include the lateral
(angular) fluxes in the transport update of the radiation fields;
we do. We each use a different number of energy groups, placed at
different neutrino energies. One difference is our neglect of the velocity-dependent terms in the
transport equation.  Buras et al. calculate the radiation field in the
comoving frame, and conclude that their inclusion decreases the net gain.
We do the calculations in the lab frame.  Hubeny and Burrows (2006)
have included the velocity terms in such a lab-frame formalism and find that
the effect {\it increases} the net gain by $\sim$10\%.  The sign depends on the
frame in which you calculate the radiation quantities.  Hence, if
we were to include the velocity terms it would lead to an increase in
the neutrino heating. However, whether that would make a qualitative
difference in our 2D 11.2-M$_{\odot}$ simulation remains to be seen.

\begin{figure*} 
\figurenum{3}
\plotone{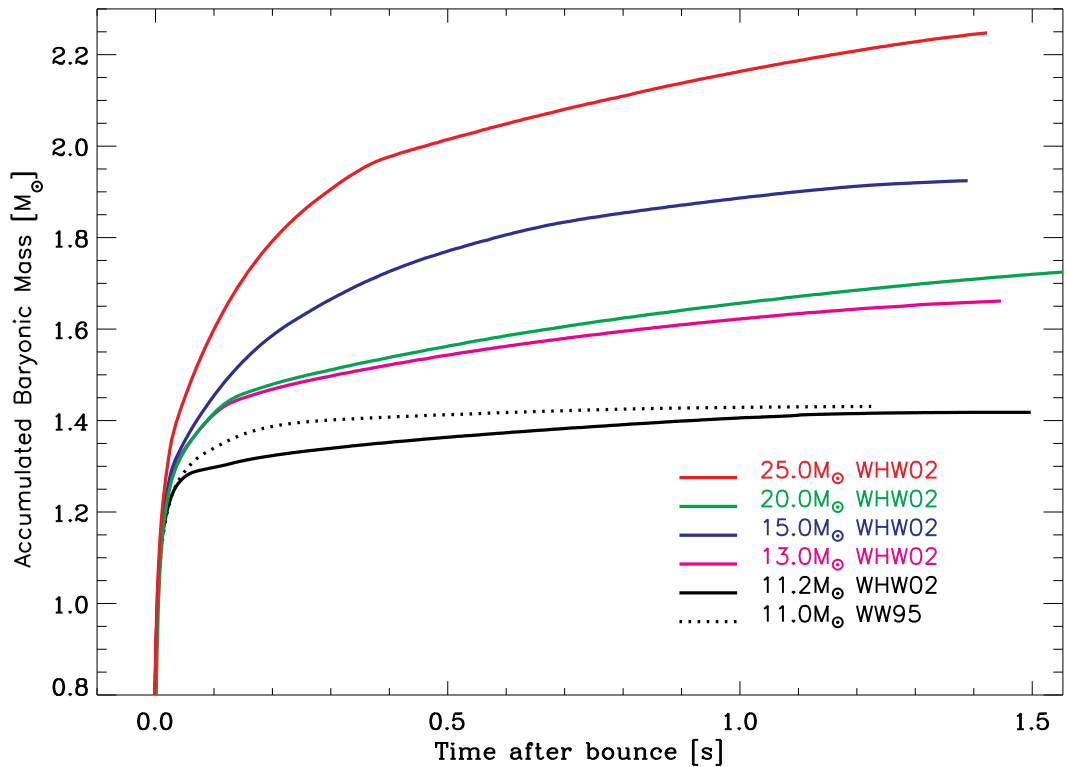}
\caption{
Time evolution after core bounce of the 
baryonic mass interior to 100\,km of the nascent protoneutron star for 
representative massive star progenitors.  See text for discussion.
}
\label{fig:3}
\end{figure*}

Figure \ref{fig:3} shows the evolution of the baryon mass accumulated in the protoneutron
star for a subset of progenitors simulated with VULCAN/2D.  When corrected
for the binding energy shed during deleptonization and neutrino cooling (Burrows \& Lattimer 1986),
these masses can be used to provide the gravitational masses of the neutron stars (or black holes)
that remain.  This plot also enables one to estimate the residual baryon mass if the 
explosion were to occur earlier, since the $\dot{M}$ evolution is fixed by the progenitor
structure (Fig. \ref{fig:1}).  One can insert on the plot a vertical line at a given time 
and read off the baryon mass remaining if the explosion of a given 
progenitor core were to occur at that time, or one can insert a horizontal line 
at a given mass to determine the time a given core must explode to leave 
that baryon mass behind.  For instance, Fig. \ref{fig:3} indicates that 
for the 20-M$_{\odot}$ progenitor of WHW02 to leave a neutron star with
a baryon mass less than 1.5 M$_{\odot}$ (equivalent to a gravitational mass of $\sim$1.35 M$_{\odot}$),
it must explode before 0.3 seconds after bounce.  Similarly, for the 25-M$_{\odot}$
model of WHW02 to leave such a neutron star, it must explode within
the first 50-100 milliseconds of bounce.  However, the 11-M$_{\odot}$
and 11.2-M$_{\odot}$ models need not explode before 1.5 seconds to leave
behind an object with a baryon mass of 1.4-M$_{\odot}$ (roughly equivalent 
to a gravitational mass of $\sim$1.28 M$_{\odot}$).  In addition, Fig. \ref{fig:3} can be used
to determine the maximum time to the explosion of a given progenitor, and for
a given nuclear equation of state, if a neutron star, and not a black hole,
is to result (ignoring any fallback).  Once the mechanism for explosion
and the actual progenitor structures have been clearly 
determined, Fig. \ref{fig:3} can also help inform any discussion
concerning the progenitor mass at which the bifurcation between neutron star
and black hole final products occurs. 

Figure \ref{fig:3} indicates that, for the more massive models 
shown, leaving behind neutron stars with baryon masses
less than 1.6 M$_{\odot}$ would require earlier explosions than we 
currently obtain.  Such ``early" explosions may require relativistic calculations,
better neutrino transport, better numerics, different progenitor models, or 3D effects.  
However, in this paper, we focus on the general, qualitative effects that emerge from
our investigations, and will not claim at this preliminary stage, 
and given the remaining compromises in our computational approach 
(see Appendix), to have arrived at final numbers.

\section{Hydrodynamic Comparisons of Different Progenitors}
\label{evolution}

\begin{figure*} 
\figurenum{4}
\plotone{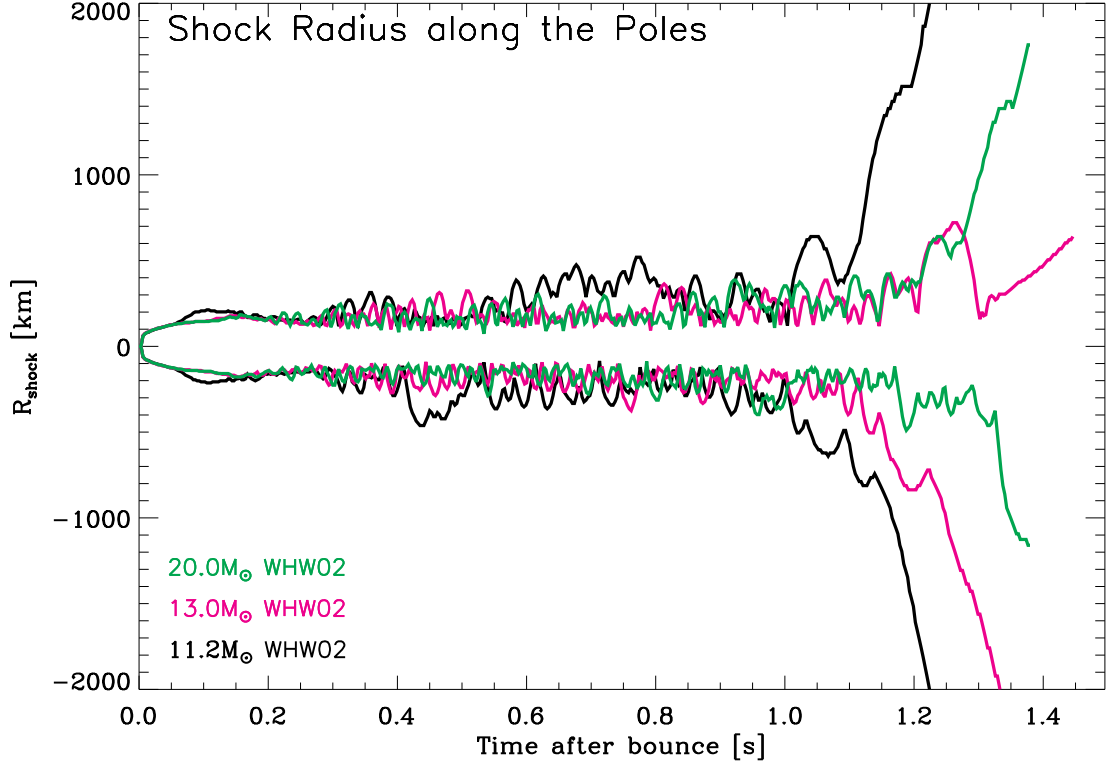}
\caption{
Time evolution of the outer shock radius (in km) along the poles for the 11.2-\mo (black), 
13-\mo (magenta), and 20-\mo (green) models of WHW02.  The radii extend from 2000 km
to -2000 km.
}
\label{fig:4}
\end{figure*}

Figure \ref{fig:4} depicts the evolution with time after bounce of the 
radial positions along the poles (in the positive and negative directions)
of the outer shock for three representative progenitor model simulations with VULCAN/2D.
The delay to explosion for all models is $\sim$1 second, with the stars with
the steepest initial density gradients (Fig. \ref{fig:1}) and 
the lowest $\dot{M}$s (Fig. \ref{fig:2}) exploding earliest.
Though the 11.2-M$_{\odot}$ model of WHW02 does not explode 
within the first $\sim$100 milliseconds of bounce, its shock 
radius is consistently larger and experiences larger excursions before 
explosion than the other models shown in Fig. \ref{fig:4}. The 
excursions during the SASI phase range from $\sim$150 km to $\sim$500 km
and can be quite dramatic and the explosions, when they eventually occur, 
are unmistakable.   After the explosion ensues, it takes only $\sim$100 ms
for the shock to reach $\sim$1000 km.  By the end of all the simulations 
performed for this study, the explosion radius has reached $\sim$6500 km 
or more along multiple directions.  

However, as Fig. \ref{fig:4} demonstrates, and Fig. \ref{fig:5},
which depicts a color map of the $\pm$polar entropy profile for four representative models,
confirms, the blasts are top-bottom asymmetric.  The SASI and the core oscillation
represent symmetry breaking and the direction of explosion depends
upon the chaotic evolution of the flow and the timing of the explosion.
We see explosions that are very unipolar (e.g., our 25-M$_{\odot}$ run) and more
top-bottom symmetric (e.g., our 11.2-M$_{\odot}$ run).  One can't 
predict ahead of time in which direction the core will explode, nor
the degree of anisotropy.  However, one can expect that the distribution
of the top-bottom asymmetries and the character of these asymmetries
can eventually be determined statistically.  Figure \ref{fig:6} portrays
snapshots of the explosion debris a few hundred milliseconds after explosion
of the 11.2-M$_{\odot}$ and 20-M$_{\odot}$ models and highlights the
different degrees of early blast asymmetry we can expect.
We have seen top-bottom asymmetries larger than that for the 20-M$_{\odot}$ model,
but none less than that of the 11.2-M$_{\odot}$ model.  Note that
we are not concluding anything about the likely progenitor dependence
of the explosion asymmetry.  On the contrary, we are merely documenting  
the diversity we see in this set of numerical realizations. 

As Fig. \ref{fig:2} demonstrates, the evolution of the mass accretion rate is  
different for the different progenitors.  Among other things, this translates into accretion
rams that are very different from model to model.  Table 1 demonstrates another
consequence: the SASI shock frequencies vary by more than a factor of 2.5
from progenitor to progenitor and there is a one-to-one relationship between the $\dot{M}$ and the SASI frequency.
We have Fourier analyzed the shock position and it is the dominant frequencies
that are listed in Table 1, along with the average accretion rates and average shock
radii during the non-linear SASI phase before explosion.  Oscillation frequencies from $\sim$30 Hz
to $\sim$80 Hz are seen and these are inversely proportional to the average radius
of the stalled shock (in Table 1, from $\sim$120 km to $\sim$250 km).  As might have been
expected, the monotonicity is with $\dot{M}$ and the shallowness of the density profile, and not the
progenitor ZAMS mass.  The oscillation periods implicit in Table 1 are approximately 
the sound-travel-times across the shocked regions. Note that this is not a statement
about the {\it growth} timescale of the SASI, which is very different and is not the 
sound-travel time (Foglizzo, Scheck, \& Janka 2006).  For the smaller average shock radii that
we obtain when the $\dot{M}$s are larger, this translates into the higher oscillation frequencies
for those models. This is in contrast to the similarity we see in the core g-mode
frequencies for the various progenitors: at a given epoch this frequency ranges
only modestly from model to model and, for our Newtonian calculations, the 
$\ell = 1$ mode sticks within $\sim$30\% of $\sim$300 Hz.

As we show in Table 1, the average shock radius during the SASI phase is smaller
for those progenitors with the highest post-bounce mass accretion rates.  The 25-M$_{\odot}$
model of WHW02 is an example of a massive star progenitor with such a high rate.
As demonstrated in Ott et al. (2006b), this model manifests not only $\ell = 1$
core oscillations, but significant $\ell =2$ core oscillations as well.  The latter
are responsible for the strong gravitational radiation signature of this published model.  Such
strong $\ell =2$ core oscillations are more easily excited if the outer SASI
shock oscillations have a strong $\ell = 2$ component as well.  Foglizzo et al. (2006)
have recently performed an analytic stability and growth-rate analysis of the SASI and
find that those models with the smallest ratio between the shock radius and the inner
core radius should experience stronger $\ell = 2$ SASI growth.  The more detailed 
2D radiation/hydrodynamic simulations reported in this paper and in Ott et al. (2006b) 
for the 25-M$_{\odot}$ model, with its more compact shock configuration, tend to bear out 
these findings.  Progenitors with larger $\dot{M}$s result in smaller shock/core radius contrasts, higher
SASI frequencies, and larger growth rates for the $\ell = 2$ modes of both
the SASI and the core oscillation.  The strong $\ell = 2$ core mode can result in prodigious gravitational
radiation signatures (Ott et al. 2006b) of the associated supernovae, and of black hole formation,
which itself may be the result of large $\dot{M}$s.

On the left in Fig. \ref{fig:7}, the evolution of the net neutrino energy 
deposition in the gain region versus time after bounce is portrayed for 
five representative progenitor models from WHW02. There is a strong dependence
of this power on $\dot{M}$.  However, these numbers are relevant only after explosion
commences and infall transitions into outflow.  Before that, the net neutrino energy
deposition for a given Lagrangean mass element changes sign as the settling mass element encounters the
inner cooling region just exterior to the neutrinospheres.  Hence, it is when
these powers start to decrease due to the reduction of the neutrino luminosities
caused by the decrease in $\dot{M}$ upon explosion that 
neutrino heating can contribute to the explosion energies,
and it does so in a transient fashion.  As Fig. \ref{fig:7}
shows, the net effect of neutrino heating from the onset of explosion,
which itself in these calculations is due predominantly to acoustic
power, is an increasing function of $\dot{M}$ and, approximately, of progenitor mass.
The largest effect of neutrino heating in this model set is for the 
25-M$_{\odot}$ model of WHW02 and amounts to an integrated value of $\sim$2$\times 10^{50}$ 
ergs by the end of the simulation near 1.4 seconds after bounce.  At this time,
the explosion is still being driven at a steady rate by acoustic power from its massive core.

On the right side of Fig. \ref{fig:7}, we provide the corresponding evolution
of the total gravitational accretion power ($\dot{M} GM/R$) for the same set
of representative models, along with the values for the 11-M$_{\odot}$ model
studied in Burrows et al. (2006).  The accretion power ranges by almost two orders
of magnitude, directly reflecting the range in mass accretion rates (Fig. \ref{fig:2}).  Most
of this power is radiated to infinity as neutrinos, without heating; only a small fraction is converted
into the mechanical energy of core oscillations, most of which damps by the emission
of sound.  Due to the chaotic and anisotropic nature
of the turbulent flow interior to the shock, it has been difficult to get an analytic handle on
the efficiency of conversion of accretion power into core pulsation energy and acoustic power. However, even the
small efficiency we find is enough to ignite and power explosion, after some delay.
Neutrino damping of core oscillation seems to be a small effect (automatically included
in our calculations), with a characteristic timescale of 5-30 seconds 
(see Ferrari et al. 2003; Miralles et al. 2004).  Artificial
damping due to low resolution and truncation errors seems to have a characteristic timescale longer than
three seconds.  However, its magnitude is difficult to gauge, given the expense of the simulations,
and must be a subject for future studies.  Most of the higher-resolution studies (both spatial and spectral) we have
performed suggest that greater resolution leads to slightly more vigorous
core oscillations and SASI, particularly when we increase the number 
of energy groups. Note that in the models described in this paper, 
the grid transitions from spherical to Cartesian at $\sim$30\,km, where 
the radial spacing is still a respectable $\sim$800\,meters.  Such resolution provides  
reasonable sampling ($\sim$5 zones per decade in density) of the steep density profile
that arises in this region at late times ($\sim$1\,s after bounce).

Figure \ref{fig:8} depicts the energy in the g-mode oscillations of the inner core
versus time after bounce for a few representative progenitor models.  There seem
to be two classes. The first, represented by the 11.2-M$_{\odot}$ and old 11-M$_{\odot}$
models with low mass accretion rates, achieve 
pulsation energies (kinetic plus potential) of only $\sim$10$^{50}$ ergs.
The second class is represented by the 20-M$_{\odot}$ model in Fig. \ref{fig:8},
for which the core achieves pulsation energies near $\sim$10$^{51}$ ergs.
For such models, the turbulence of the SASI and the compactness
of the shock are much greater (Table 1), and the mass accretion rates are much larger.
After explosion commences, the inner cores reach quasi-steady states 
in which the fraction of the gravitational accretion energy 
channeled into mechanical energy roughly balances the acoustic losses.  
This happens at an acoustic power of very approximately $\sim$0.5$\times 10^{51}$ ergs s$^{-1}$.  
The available accretion energy subsides with explosion (Fig. \ref{fig:7})
and so the energy stored in the g-mode within 100-200 milliseconds
of the onset of explosion may be a measure of the total energy available to be pumped into
the supernova ``nebula" acoustically.  At a loss rate of $\sim$0.5$\times 10^{51}$ ergs s$^{-1}$,
the low-accretion-rate progenitors would achieve explosion energies of a few $\times 10^{50}$ ergs
within hundreds of milliseconds and the high-accretion-rate progenitors would achieve explosion energies of
$\sim$10$^{51}$ within seconds, the time it would take their larger core oscillations
to discharge acoustically\footnote{The old 11-M$_{\odot}$ model of WW95
studied in Burrows et al. (2006) exploded earlier because its inhibiting mass accretion rate
was very small and because the outer boundary radius was put at too small a value (3400 km).
The new models all have a larger outer radii of $\sim$6500 km.  
For the 11-M$_{\odot}$ model of WW95, Burrows et al. (2006) found that the 
acoustic pumping lasted $\sim$400 milliseconds.}.  General relativity will increase 
the core frequencies, and, hence, the core acoustic power, with the result that
the explosion might occur earlier and, perhaps, more energetically. The dependence
on the nuclear EOS is more subtle and has yet to be studied. 

Whether the bifurcation into two classes is abrupt, or whether there is in reality 
a continuum from lower to higher core pulsation energies, remains to be seen.
However, progenitors with larger mass accretion rates seem to achieve larger
core pulsation energies, with the suggestion that they can explode more energetically.
As the left panel of Fig. \ref{fig:7} also suggests, the neutrino contribution
to the explosion energy is expected to be larger for progenitors with higher 
$\dot{M}$s at explosion, or for progenitors that for some reason explode earlier (all else being equal).
Unfortunately, due to difficulties at the outer computational boundary, which must handle 
simultaneous infall and explosion, and/or convergence problems in the neutrino-matter coupling 
in the inner core material residing in the transition region from Cartesian 
to spherical gridding when it becomes very violently pulsational 
and the mass density gradients steepen precipitously,
we are not yet able to evolve our models beyond $\sim$1.5 seconds after bounce.  The code crashes.  
Hence, we do not quote total explosion energies.  However, the
acoustic power being pumped into the explosion ($\sim$10$^{50}$ erg s$^{-1}$ 
to $\sim$10$^{51}$ erg s$^{-1}$) and the core oscillation energy
ultimately available to the supernova by acoustic discharge 
($\sim$10$^{50}$ ergs to $\sim$10$^{51}$ ergs) give us zeroth-order 
estimates of the systematics and values of the final explosion 
energies.  From our results, the initial supernova explosion energy seems to be 
an increasing function of progenitor mass, when correction is made
for the slight non-monotonicities noted in Fig.~\ref{fig:1}. Whether this conclusion
survives will be contingent upon future detailed investigations, using a variety 
of techniques and codes. We note that Hamuy (2003) has inferred   
from observations of supernova explosions that explosion energies
might in fact span a wide range of values.

\section{Entropy and Electron Fraction of the Ejecta}
\label{entropy}

The explosions we see involve ejecta with distributions of entropies
and electron fractions (Y$_e$).  If the ejecta entropies achieve
values in the hundreds, it has been shown that r-process nucleosynthesis
becomes more viable (Woosley et al. 1994).  We have assembled histograms of the
amount of mass in the escaping fraction in the various entropy and Y$_e$
bins.  However, since the explosions have not run to completion (despite
the fact that the blasts have reached 6500 km and the simulations 
have been performed to $\sim$1.5 seconds), we do not have final histograms
for any of our simulations.  Nevertheless, the numerical data we do have 
are intriguing and we present them in Fig. \ref{fig:9} 
for the 11.2-M$_{\odot}$ and 20-M$_{\odot}$ runs.  The heights give
the logarithm of the total mass in the Y$_e$ and entropy (actually log(entropy))
bins and are not differentials.  For the 11.2-M$_{\odot}$ and 20-M$_{\odot}$ runs, we find
that the total masses ejected above entropies of 100 per baryon per Boltzmann's
constant are 2.15$\times 10^{-4}$ M$_{\odot}$ and 1.1$\times 10^{-5}$ M$_{\odot}$,
respectively, while the total masses ejected above entropies of 300 per baryon per Boltzmann's
constant are 1.25$\times 10^{-4}$ M$_{\odot}$ and 0.0 M$_{\odot}$,
respectively.  The total masses ejected at any entropy are 0.0191 M$_{\odot}$
and 0.0041 M$_{\odot}$ for the 11.2-M$_{\odot}$ and 20-M$_{\odot}$ models,
respectively.  For core-collapse supernovae
to be the site of the r-process, each must eject on average $10^{-4}$ to $10^{-5}$
M$_{\odot}$ of r-process elements (Woosley \& Hoffman 1992; Woosley et al. 1994; Hoffman et al. 1996; 
Thompson, Burrows, \& Meyer 2001). The r-process yield in a parcel of matter
varies with entropy, Y$_e$, and expansion time, but can be around 10\% by mass for the highest entropies
and the ``long" expansion times (hundreds of milliseconds) we find.  Most of the rest
of these inner ejecta will emerge as $\alpha$-particles.  The iron peak would be produced as the 
shock encounters and traverses the oxygen shell on timescales typically longer 
than those of these simulations.  Note that an upper bound of 0.5 to the ejecta Y$_e$
was inadvertantly imposed on these runs. Since Y$_e$ was not allowed to exceed 0.5, 
the potential effects of $\nu_e$ and $\bar{\nu}_e$ absorption for 
Y$_e$s above 0.5 and/or in enabling the $\nu$-p and rp processes 
(Fr\"olich et al. 2005ab,2006; Pruet et al. 2005,2006) were not 
properly incorporated.  Nevertheless, the purpose of Fig. \ref{fig:9} is to
demonstrate that high entropies are achieved and this 
conclusion is not effected by the ``$\le 0.5$" constraint.
Furthermore, our necessary use of an MGFLD algorithm, instead 
of full Boltzmann transport, for these multi-D runs should in itself 
and in any case produce less reliable values for the ejecta Y$_e$s.  

The histogram on the left-hand-side of Fig. \ref{fig:9},
depicting the results for the 11.2-M$_{\odot}$ model with a significant
amount of ejecta above entropies of 300,  suggests that the r-process yield
of that model is in the middle of the desired range.  We have yet to post-process
our ejecta with detailed nucleosynthesis codes, and so our results are
at best preliminary.  However, ours are the first consistent supernova calculations that both
explode and eject matter with true r-process potential.

The large entropies achieved in the acoustic/core-oscillation mechanism are in part a consequence
of the late explosion in lower-density matter and of the compound effects of multiple
shocks originating from the multiple sound pulses.  However, neutrino heating of the matter
made thinner by acoustic driving is a factor as well. Had the supernova explosion,
actually a wind, been driven exclusively by neutrinos, they would have
been responsible for the density profile as well.  The entropies, densities, temperatures,
and Y$_e$s of this wind would all have been determined by the driving neutrino luminosity 
and would have been inadequate to achieve the high entropies necessary for 
r-process conditions (Thompson, Burrows, \& Meyer 2001). 
However, since both acoustic and neutrino driving are simultaneously operative,
the neutrinos can deposit energy in material already made thinner by the acoustic
effects, resulting in higher entropies than can be achieved by neutrinos alone. 
Analysis of the contributions of these different agents to entropization is made next to impossible 
by the multiple and chaotic reflections and reverberations of sound waves and shocks off the walls
of the cavity into which the lion's share of core acoustic energy is being pumped.

There is simultaneous explosion and accretion, enabled by the symmetry breaking.
Given symmetry breaking, and without rotation, the explosion naturally generates a cocoon which 
roughly collimates the outflow. As this cavity is filled with acoustic 
power, it expands outward, wrapped by the infalling matter being diverted 
to the ``back" side, most of which, during the earlier stages of explosion,
is still accreted in sheets/funnels onto the protoneutron star (see Burrows et al. 2006 and Fig. \ref{fig:6}).  
The sound speeds in the exploding cavity are much larger than the initial
speed of the outer shock/explosion wave and the matter is very radiation-dominated
(as the large entropies in evidence in Fig. \ref{fig:9} would imply).
The relatively slow speed (compared with the speeds of the multiple shocks emanating from the core)
of the expansion of the blast as it works its way out, 
deflecting the accreting matter on that one side as it moves, allows the 
entropy of the cavity to accumulate and grow.  Had the cavity expanded on dynamical times,
the entropies achieved would have been much lower.
In this way, high entropies are achieved.

Since we have yet to follow our simulation explosions to completion, the systematics with 
progenitor mass of the ejecta entropy, and hence perhaps of the r-process yields,
is not obvious.  Nevertheless, Fig. \ref{fig:9} is suggestive.

\section{Anisotropic Wind and Anisotropic Neutrino Flux}
\label{aniso-wind}

The explosions we see resemble strong anisotropic winds (see \S\ref{discussion}).
A spherical wind imparts no net momentum to the residue; an asymmetric wind
imparts a kick and ``ablation" force on the accretion streams and core.
The recoil implied is a purely hydrodynamical mechanism, whatever the
agency of explosion (be it neutrinos or sound), and has two results.
First, the recoil due to the anisotropic wind pushes the accretion streams
to the opposite side, making the accretion very anisotropic.  A fraction
of the gravitational energy of accretion is used to continue to excite the inner
core g-mode oscillation.  Because the accretion funnels are supersonic,
the coupling to the core is non-linear.  Importantly, the oscillation
of the core can not do work back on the exciting accretion stream(s) that
would otherwise damp the core oscillation; any work done is accreted back. Hence, the analogy
with the swing which requires a resonance or near resonance to achieve
significant amplitude is not germane.  A steady stream onto the core
can continue to power the periodic core oscillation, even though there
is no intrinsic periodicity to the accretion.  The accretion funnel
does have a width, which like a rock hitting a pond has associated with
it a range of characteristic sizes (read wavelengths/wavenumbers).  Due to the
dispersion relation of gravity waves between wavenumber and frequency, a whole period spectrum
of ripples is generated which contains the period of the $\ell = 1$ core g-mode
(as well as those of many of the higher-$\ell$ core g-modes).

Second, the recoil provides a kick to the residual core, the protoneutron star,
and this recoil may be the origin of pulsar proper motions.  The anisotropic/top-bottom
explosion acts like rocket exhaust and momentum conservation does the rest (Burrows et al. 2007, in preparation).
The magnitude of the effect can be approximated as follows: the recoil
force is equal to $\sin(\alpha){v}\dot{M_e}$, where $\sin(\alpha)$ is the
average ``anisotropy parameter," $v$ is the characteristic wind velocity,
and $\dot{M_e}$ is the wind mass loss rate.  The ``anisotropy parameter"
is defined by this expression and is a dimensionless measure of the 
dipole moment of the momentum density of the ejecta.  Its product with 
the magnitude of the ejecta velocity yields the net specific recoil momentum.  
For isotropic ejecta, $\sin(\alpha)$ is zero.  The power poured
into the supernova ``nebula" by the wind is $1/2 \dot{M_e}v^2$.  Integrating
both these quantities over time gives the net impulse and explosion energy ($E$),
respectively.  The impulse is equal to the residue mass ($M_{pn}$) times the kick velocity ($v_k$).
Taking the ratio of these two expressions results in a formula for the kick velocity:

\begin{equation}
v_k = 2E/(M_{pn}v)\sin(\alpha) \, .
\end{equation}

If we assume that the scale of $v$ is set by a sound speed ($\sim$30,000-100,000 km s$^{-1}$), we derive that
$v_k \sim 1000 (E/10^{51} {\rm ergs}) \sin(\alpha)$ km s$^{-1}$. The average
observed/inferred kick speed is 300-400 km s$^{-1}$ (Taylor \& Cordes
1993; Lyne \& Lorimer 1994), so this number is tantalizing.
The anisotropy parameter, $\sin(\alpha)$, can be large, but depends on the stochasticity of the flow.
This formula works whether the explosion is driven by neutrinos or sound, and depends
only on the wind-like character of the asymmetric explosion and simple momentum conservation.  Note that,
all else being equal, we would expect larger kicks for larger explosion
energies.  Whether all else is in fact ``equal" remains to be seen, and this correlation
may be only statistical.  We would also expect that the kicked protoneutron star and the {\it inner}
ejecta would move in opposite directions. This is a firm prediction of the model (see also Scheck et al. 2004,2006).
The correlation observed by Wang, Lai, \& Han (2006) between the spin axis and the kick direction
would naturally follow in our kick mechanism, as long as the
rotation axis sets the axis along which the SASI and the core oscillation break spherical symmetry.
This seems plausible, but whether even slight rotation, that
otherwise has only modest dynamical effect, can enforce this axis most of the time
will require 3D simulations to determine.

\begin{figure*} 
\figurenum{5}
\plottwo{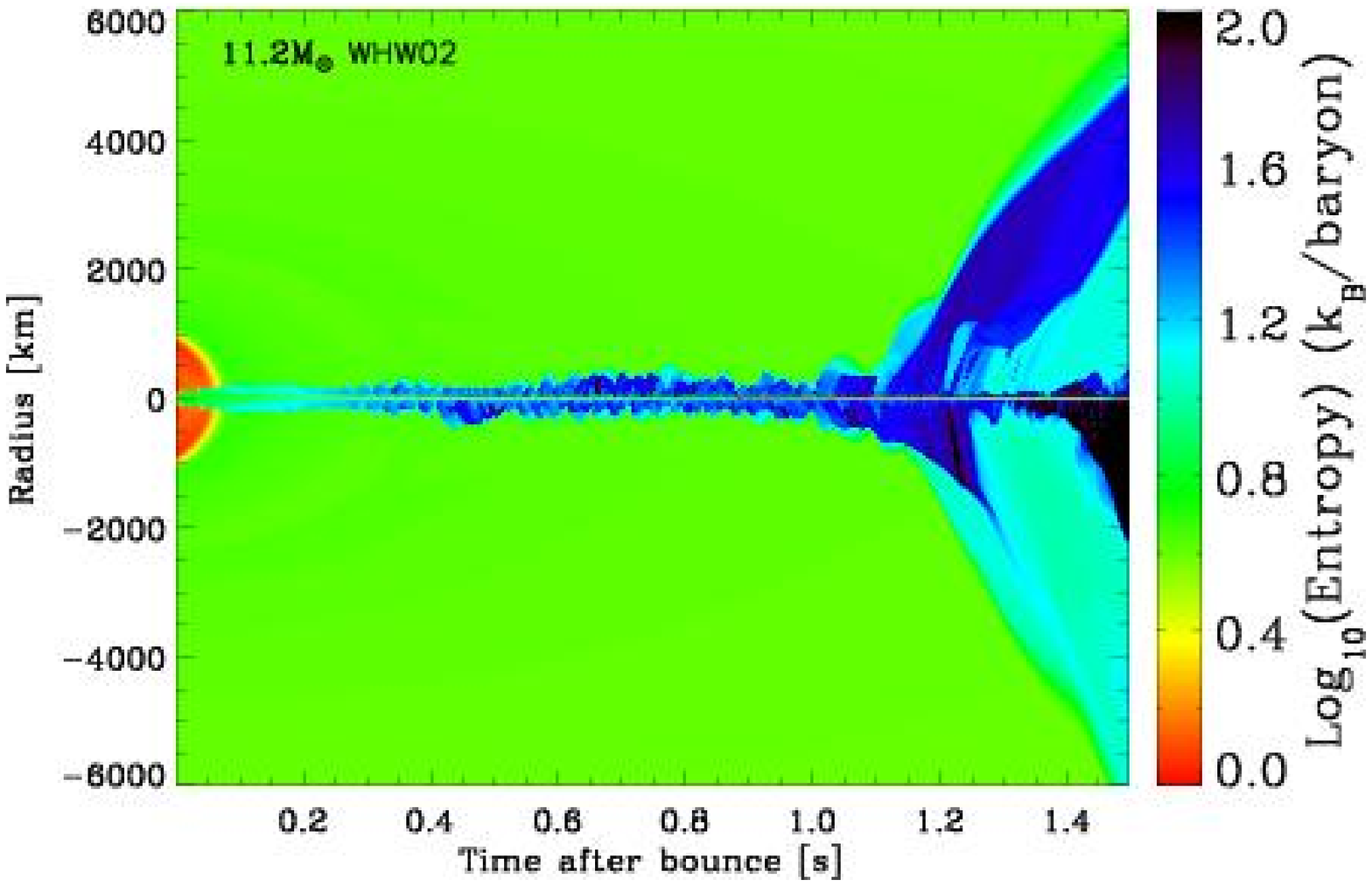}{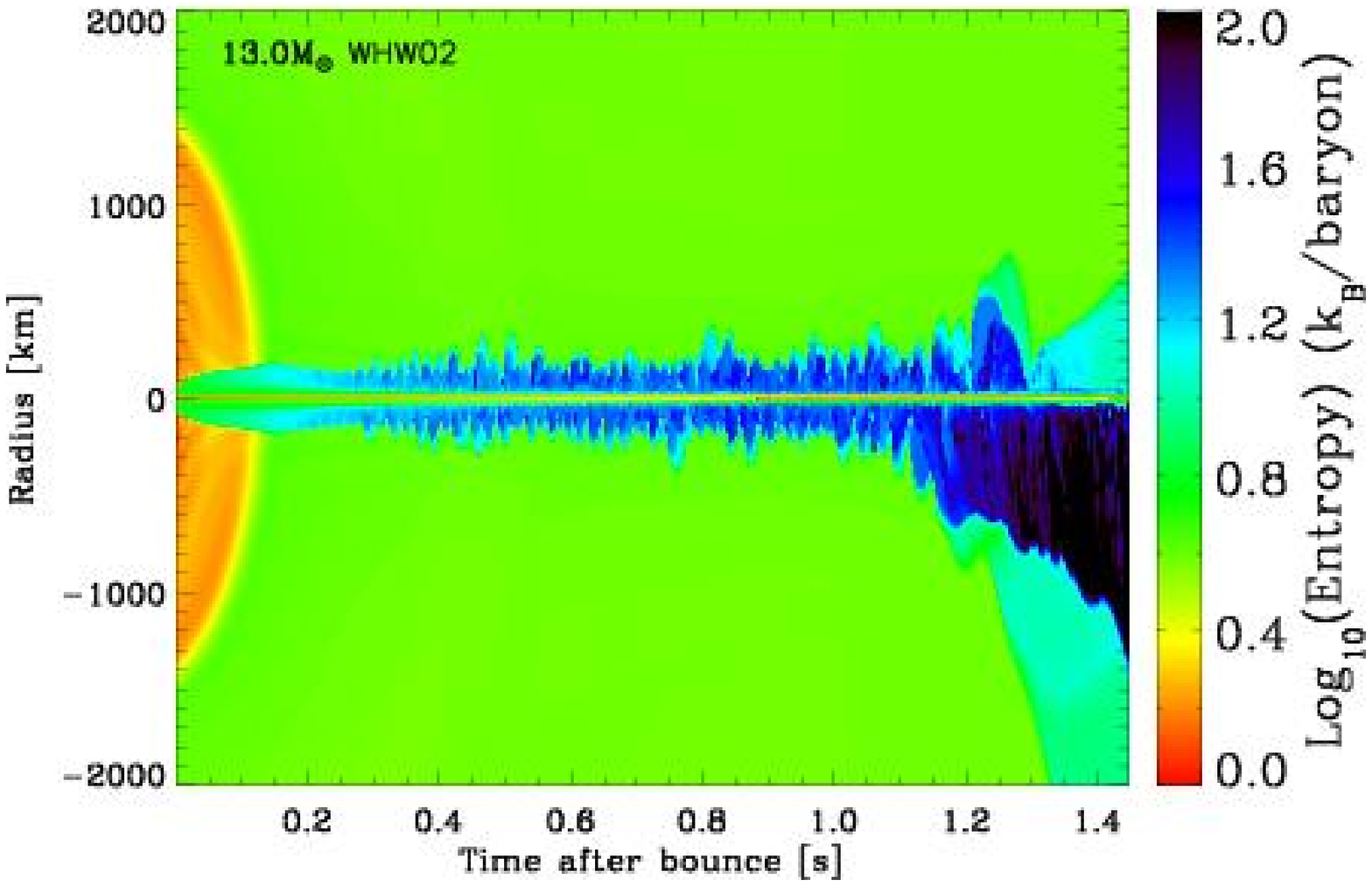} 
\plottwo{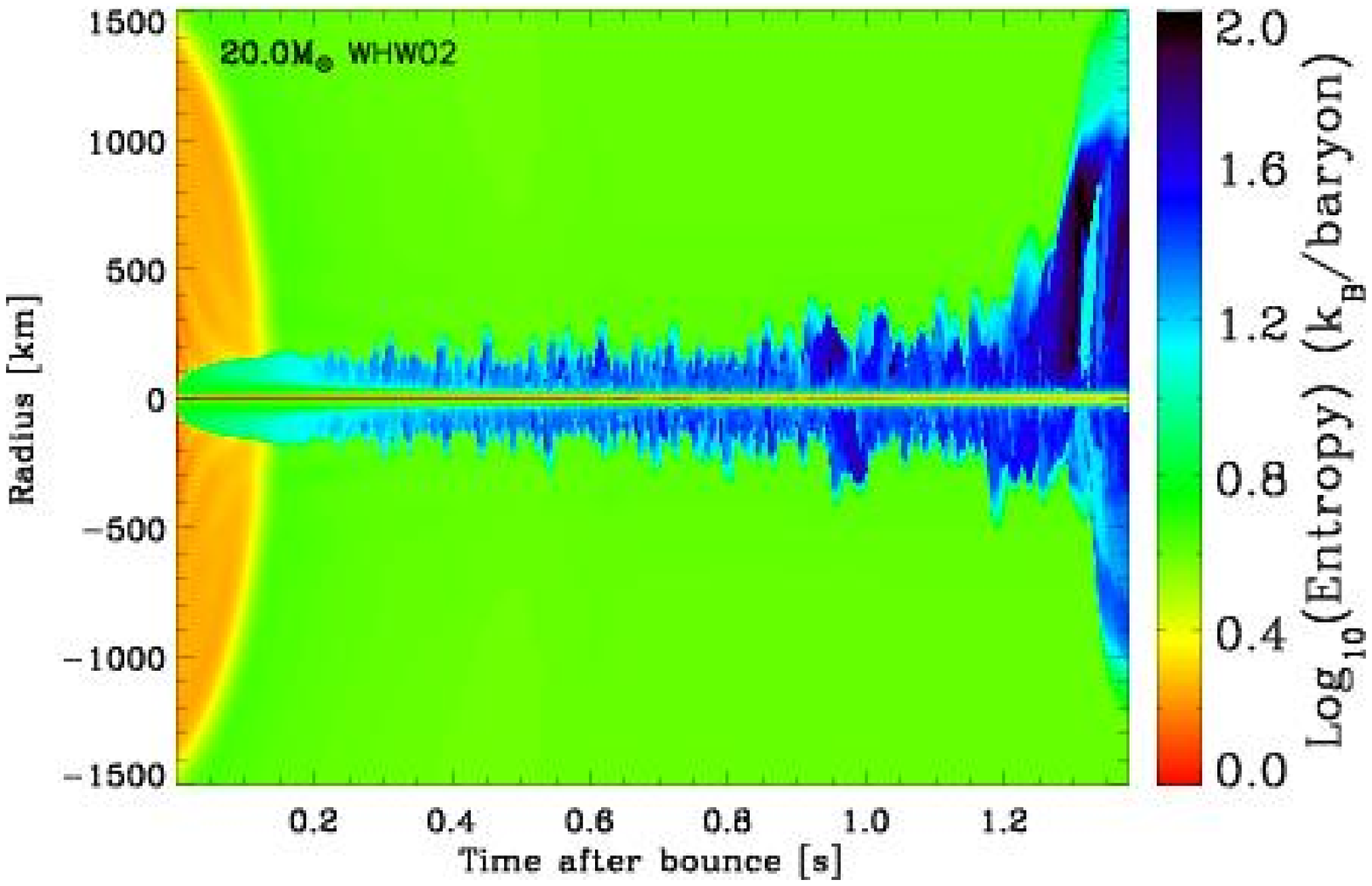}{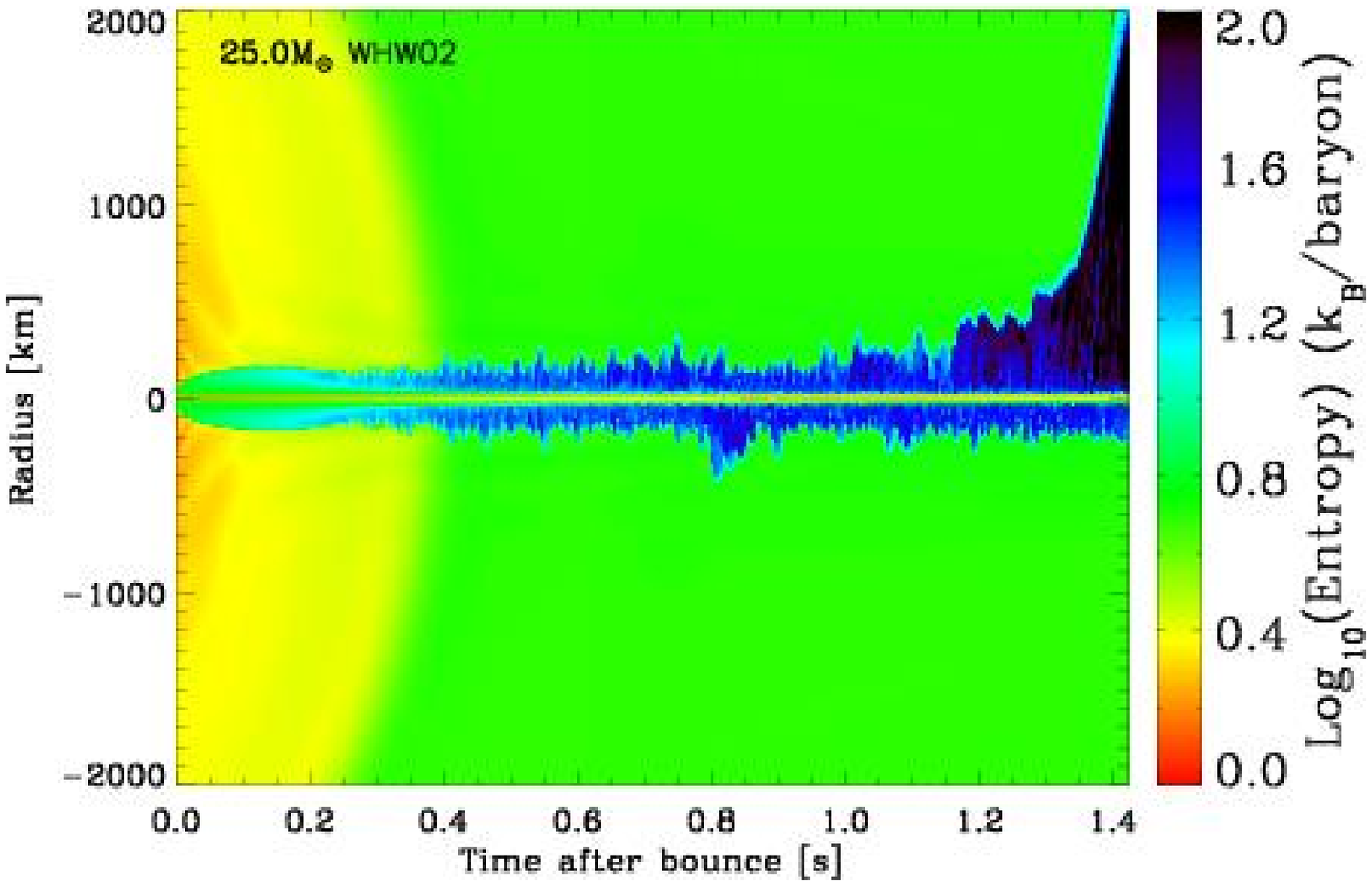}
\caption{
Time evolution of the entropy profiles along the poles of the 11.2-\mo (top-left panel),
13-\mo (top-right panel), 20-\mo (bottom-left panel), and 25-\mo (bottom-right panel)
models of WHW02.  The positions of the shocks are clearly indicated by the abrupt
transition from the green color (low entropy) of the infalling material.  
Color bars indicating the values of the logarithm of the entropy (per baryon per Boltzmann
constant) are provided on the right-hand-sides of each panel and go from red (entropy $\sim$ 1)
to purple (entropy $\ge$ 100 per baryon per Boltzmann's constant).
}
\label{fig:5}
\end{figure*}

\begin{figure*}
\figurenum{6}
\plottwo{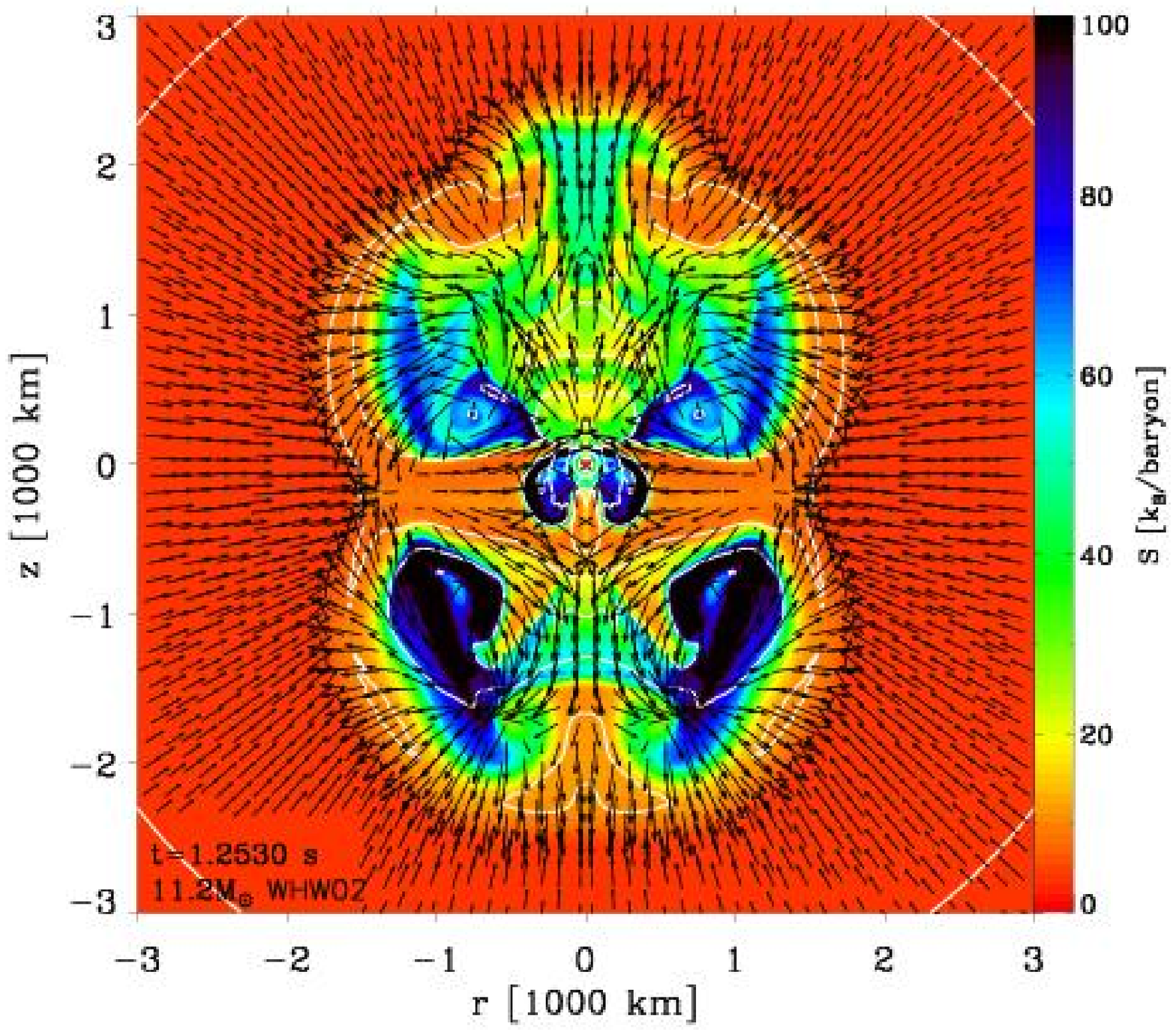}{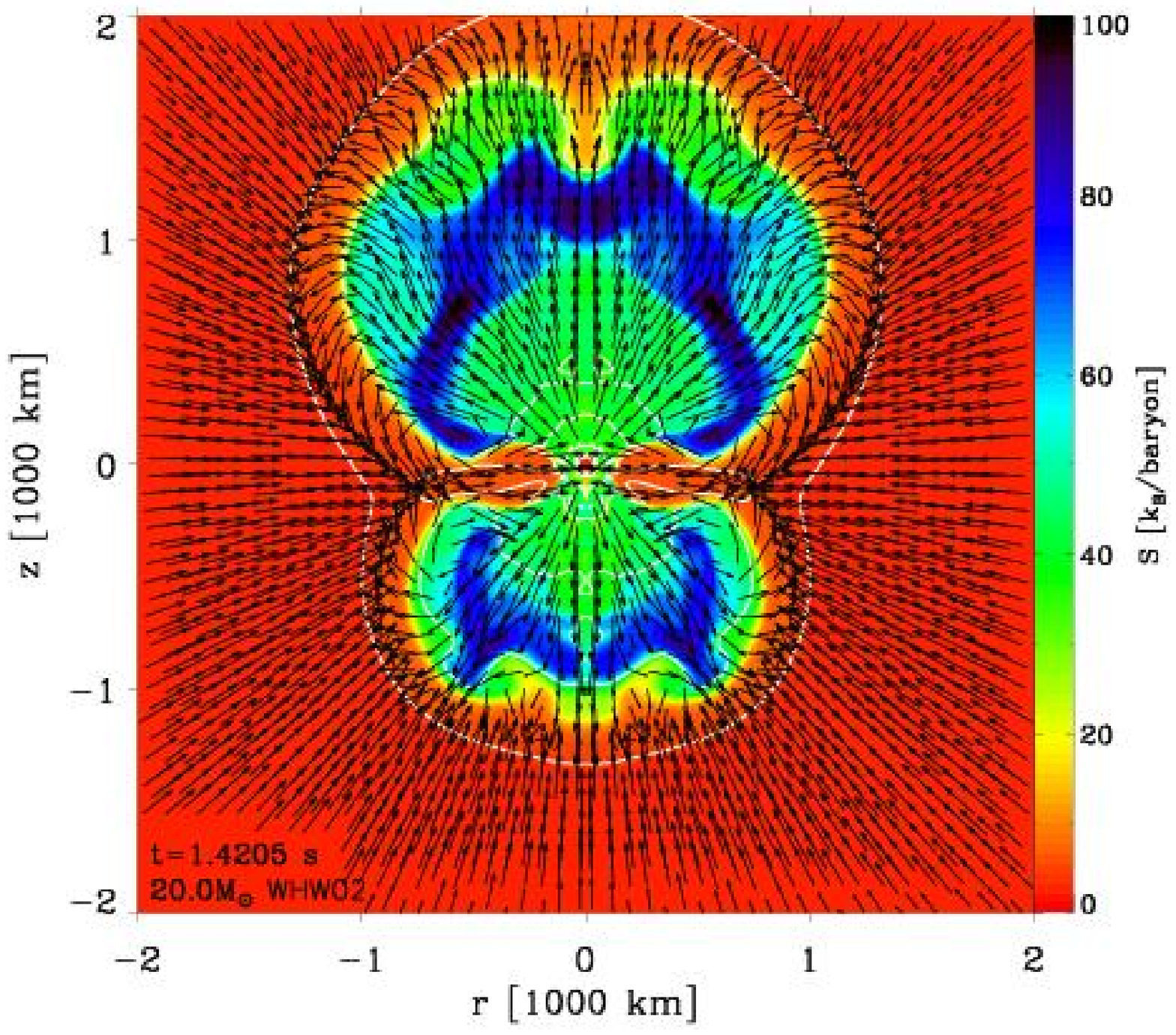}
\caption{
Entropy colormap for the 11.2-\mo (left) and 20-\mo (right) models of WHW02.
Times after bounce are indicated in the lower left hand corner of each panel.
The vector length has been saturated at a value of 10000\kms, relevant only for the
infalling matter exterior to the shock.
}
\label{fig:6}
\end{figure*}

\begin{figure*} 
\figurenum{7}
\plottwo{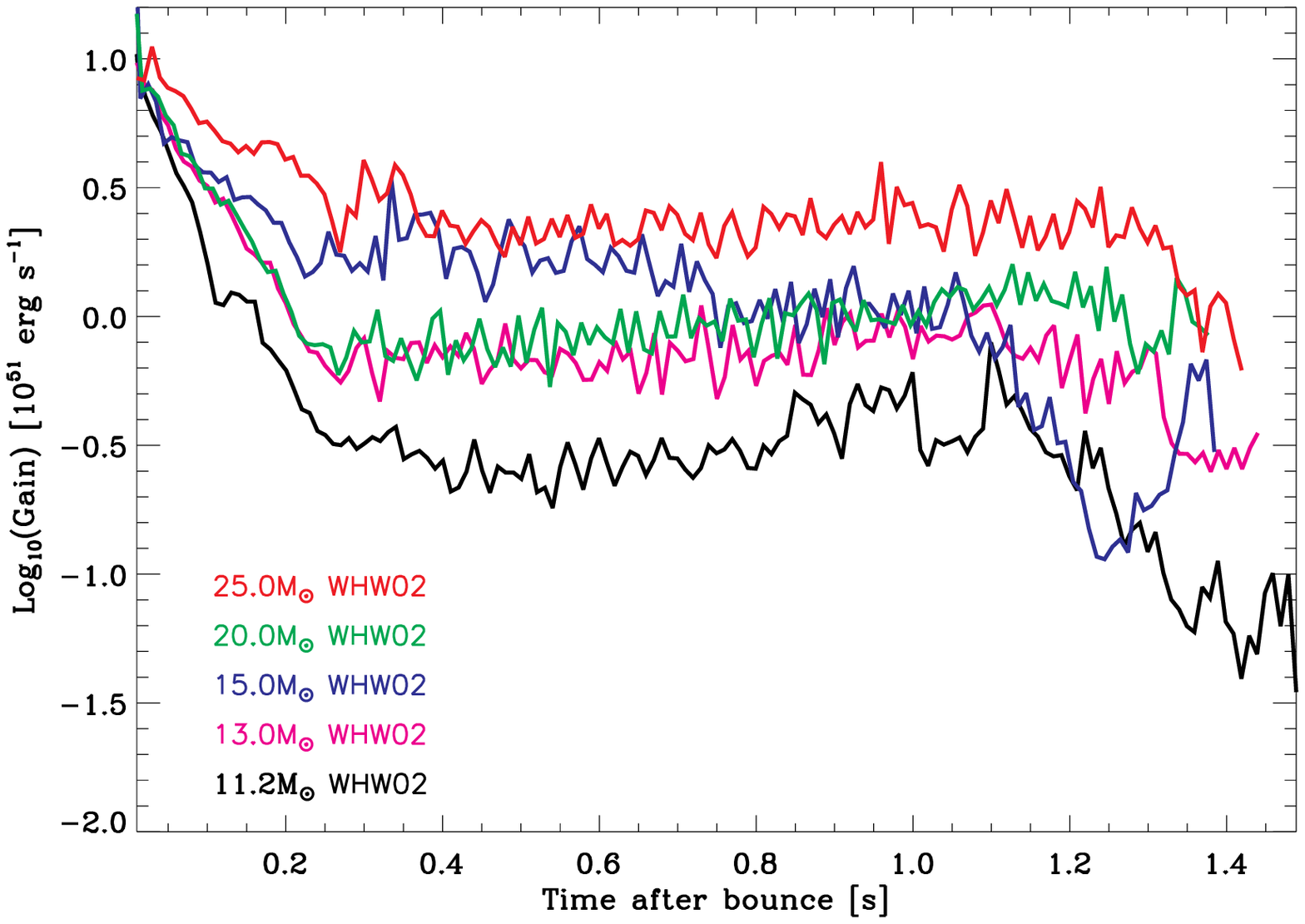}{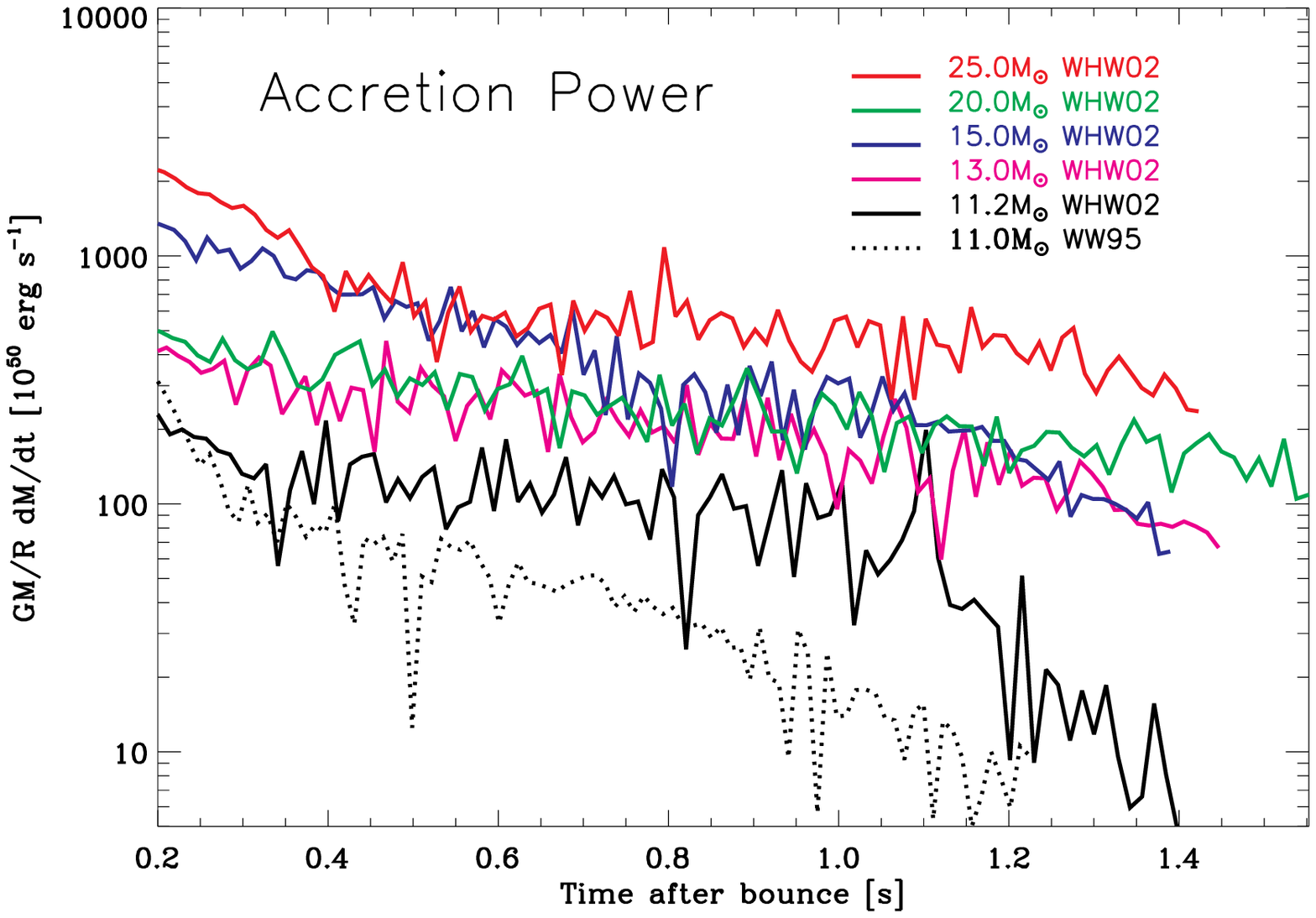}
\caption{
{\it Left}: Time evolution after bounce of the integrated net energy deposition 
due to neutrino absorption (in units of 10$^{51}$ 
erg s$^{-1}$) in the gain region for representative
WHW02 models (11.2\,\mo: black; 13\,\mo: magenta; 15\,\mo: blue; 
20\,\mo: green; 25\,\mo: red).
{\it Right}: Same as left, but for the accretion power, defined as
$\dot{M}GM/R$, where $G$ is the gravitational constant,
$M$ the mass interior to the spherical radius $R=$30\,km,
and $\dot{M}$ is the infall mass accretion rate at 30 km.
Also included is the 11-\mo model (dotted line) of WW95
calculated in Burrows et al. (2006).
}
\label{fig:7}
\end{figure*}

\begin{figure*} 
\figurenum{8}
\plotone{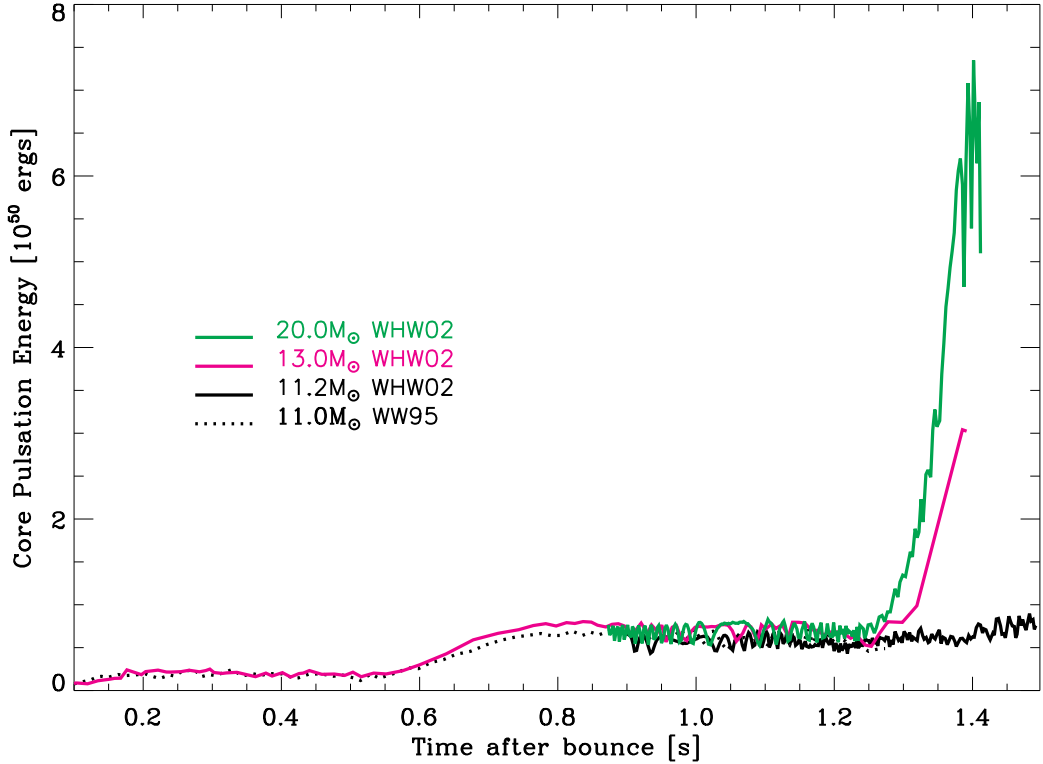}
\caption{
The time evolution after 0.1 seconds after bounce of the total pulsation energy
(kinetic plus internal plus gravitational, in units of 10$^{50}$ ergs)
of the inner cores for simulations for a representative subset of progenitor models.  
Time is given in seconds after bounce.  Note that this is not the total energy of the 
explosion at a given time.  However, since non-sonic damping processes seem weak,
the core oscillation will discharge sonically into the outer expanding ``nebula" 
and this oscillation energy will eventually be available to the explosion.  Since our calculations
halted before this phase was well underway, but after the onset of explosion,
we don't yet have a good estimate of the final explosion energy for these simulations.  
Curiously, progenitors with the largest $\dot{M}$s seem to have the largest 
total pulsation energies. See text for a discussion.
}
\label{fig:8}
\end{figure*}

\begin{figure*} 
\figurenum{9}
\plottwo{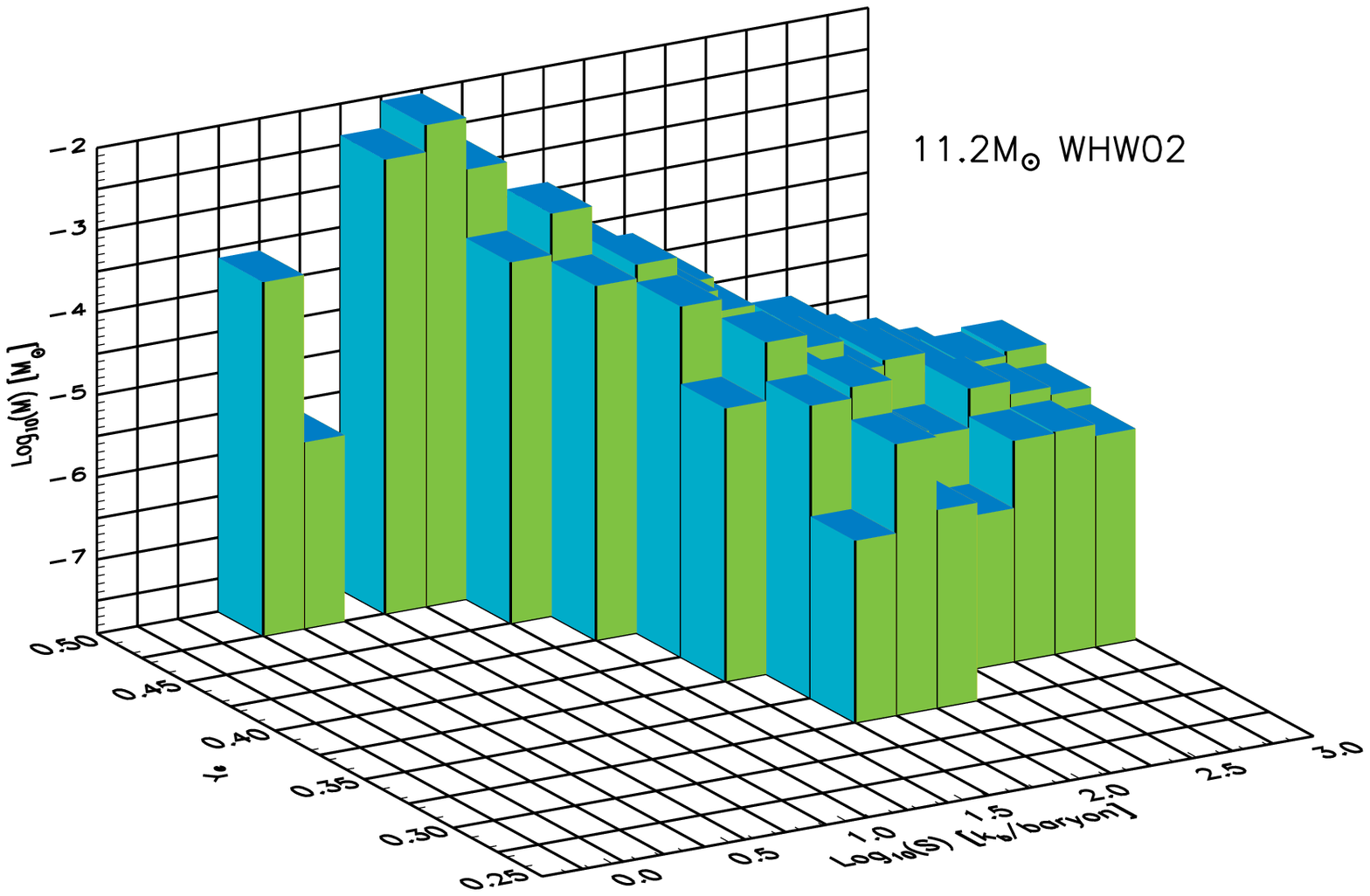}{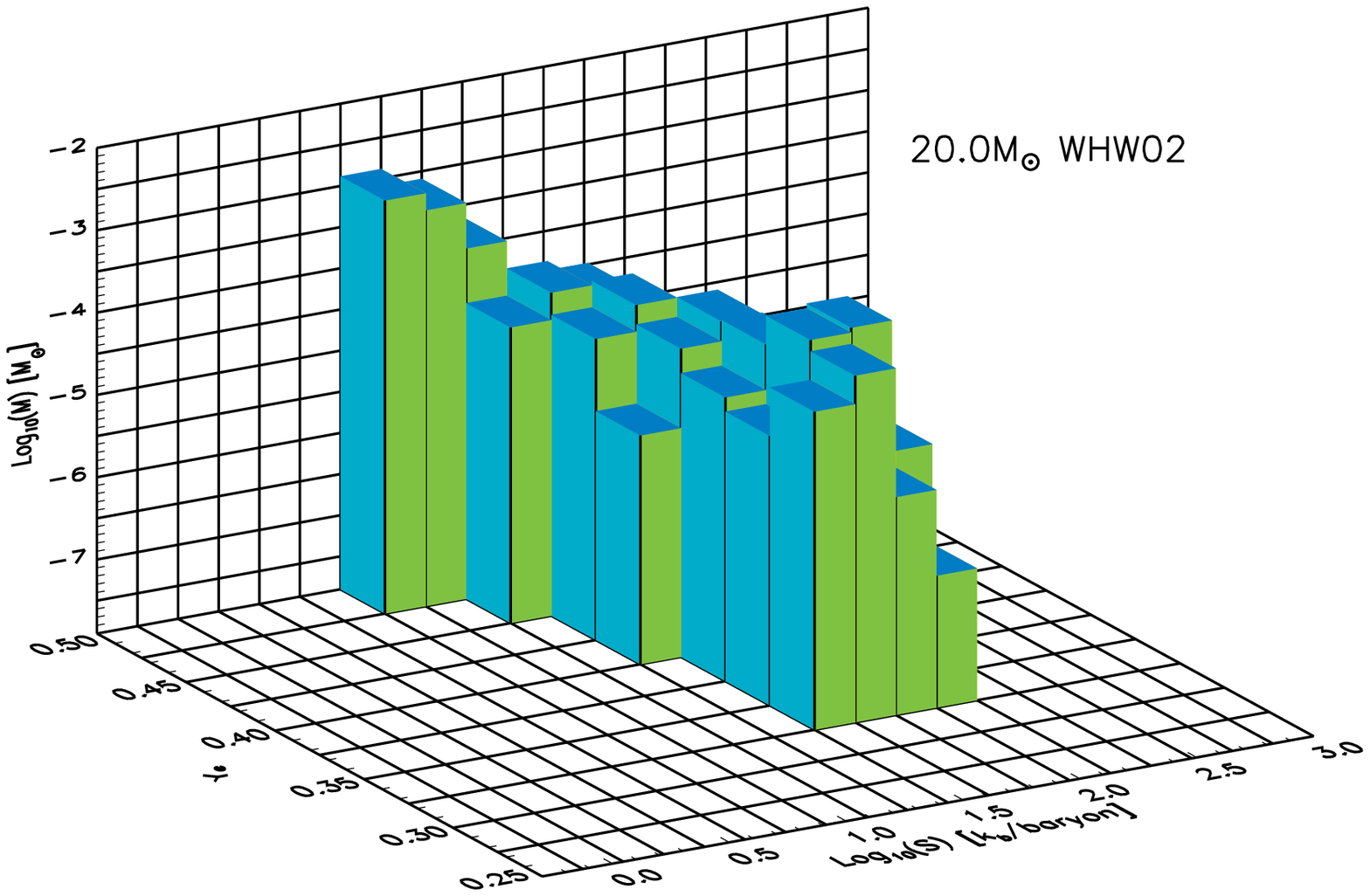}
\caption{
Histograms for the 11.2-\mo (left) and 20-\mo (right) models of WHW02
of the distribution of mass versus entropy and electron fraction (Y$_e$)
in the explosion ejecta.  The ejecta are defined as those 
parcels of matter having a positive total energy at 1500\,ms 
after core bounce. The heights of the columns are the 
actual masses (actually, logarithm of the mass, in \mo). 
For the 11.2-\mo model, the total mass ejected at the end of the
simulation, i.e. at 1.49\,s, is 0.0191\mo, while
the total mass above an entropy of 100\,k$_{\rm B}$/baryon is 
2.15$\times$10$^{-4}$\mo and above
300\,k$_{\rm B}$/baryon is 1.25$\times$10$^{-4}$\mo.
For the 20.0-\mo model, the total mass ejected at the end of the
simulation, i.e. at 1.4\,s, is 0.0041\mo, while
the total mass above an entropy of 100\,k$_{\rm B}$/baryon is
1.1$\times$10$^{-5}$\mo and above 300\,k$_{\rm B}$/baryon is 0.0 \mo.
}
\label{fig:9}
\end{figure*}

\begin{figure*}
\figurenum{10}
\plotone{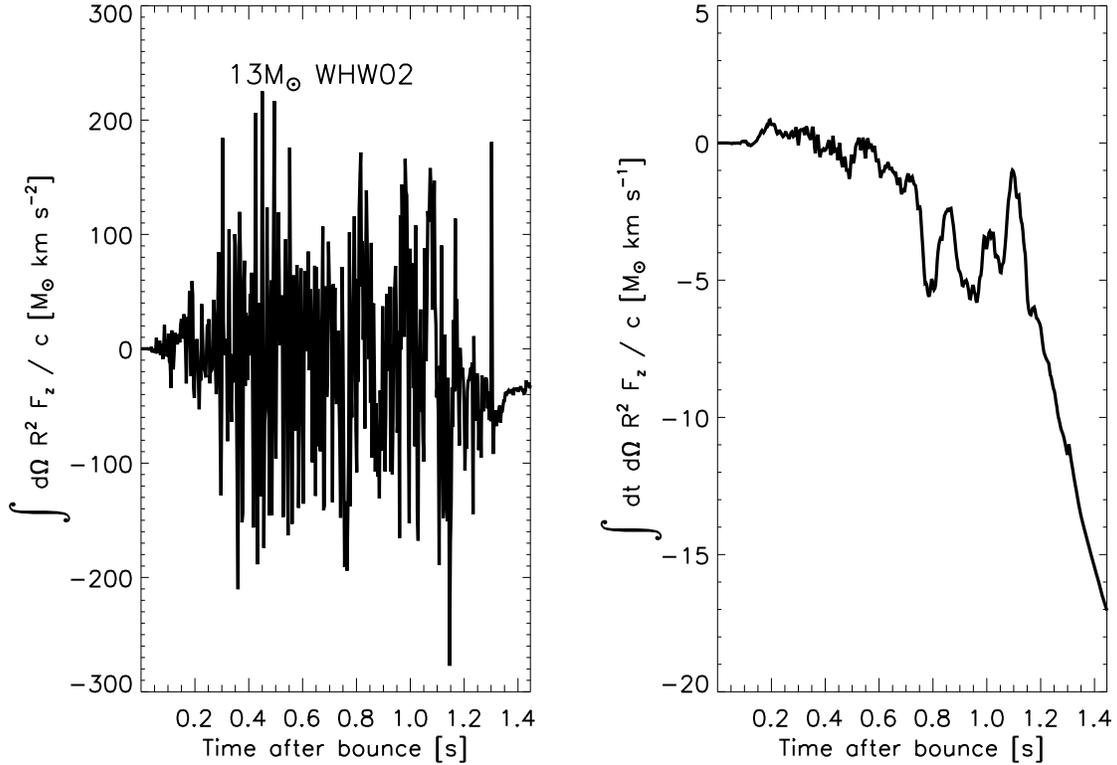}
\caption{
{\it Left:} Angle-averaged momentum of the emergent neutrinos as
a function of the time after bounce for the 13-M$_{\odot}$ model of WHW02.
This, with a negative sign, is the impulse to the protoneutron star 
due to neutrino recoil effects. Hence, the response of the core 
is in the ``positive" direction.
{\it Right:} Time integral of the instantaneous momentum shown in the left panel
as a function of time after bounce. Given the $\sim$1.6\,\mo neutron star
formed in this collapse after $\sim$1\,s, the kick imparted through the
anisotropy of the neutrino luminosity is on the order of
$\sim$10\,km\,s$^{-1}$ by the end of this calculation, 
but will likely keep accumulating.
}
\label{fig:10}
\end{figure*}

\begin{figure*}
\figurenum{11}
\plottwo{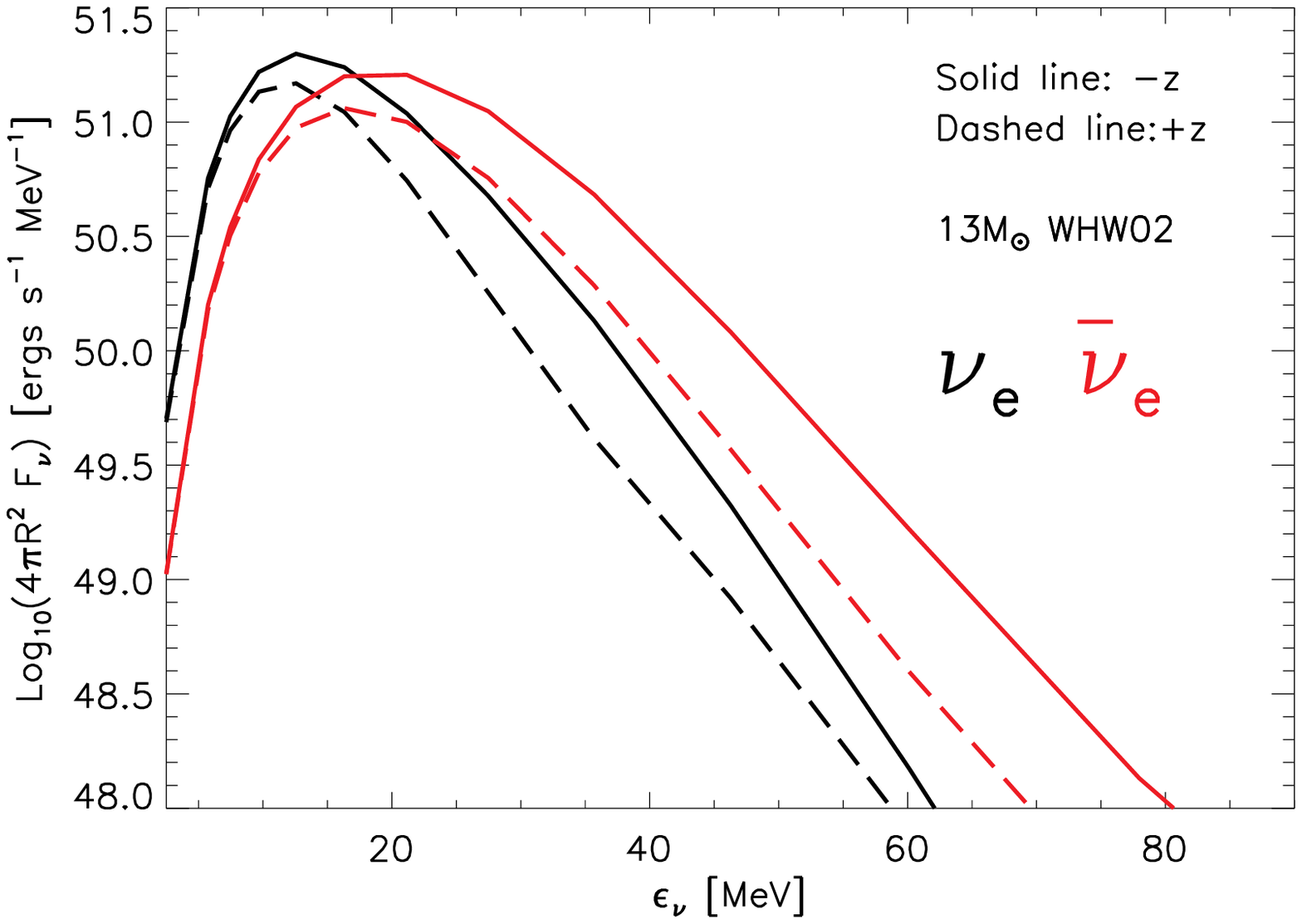}{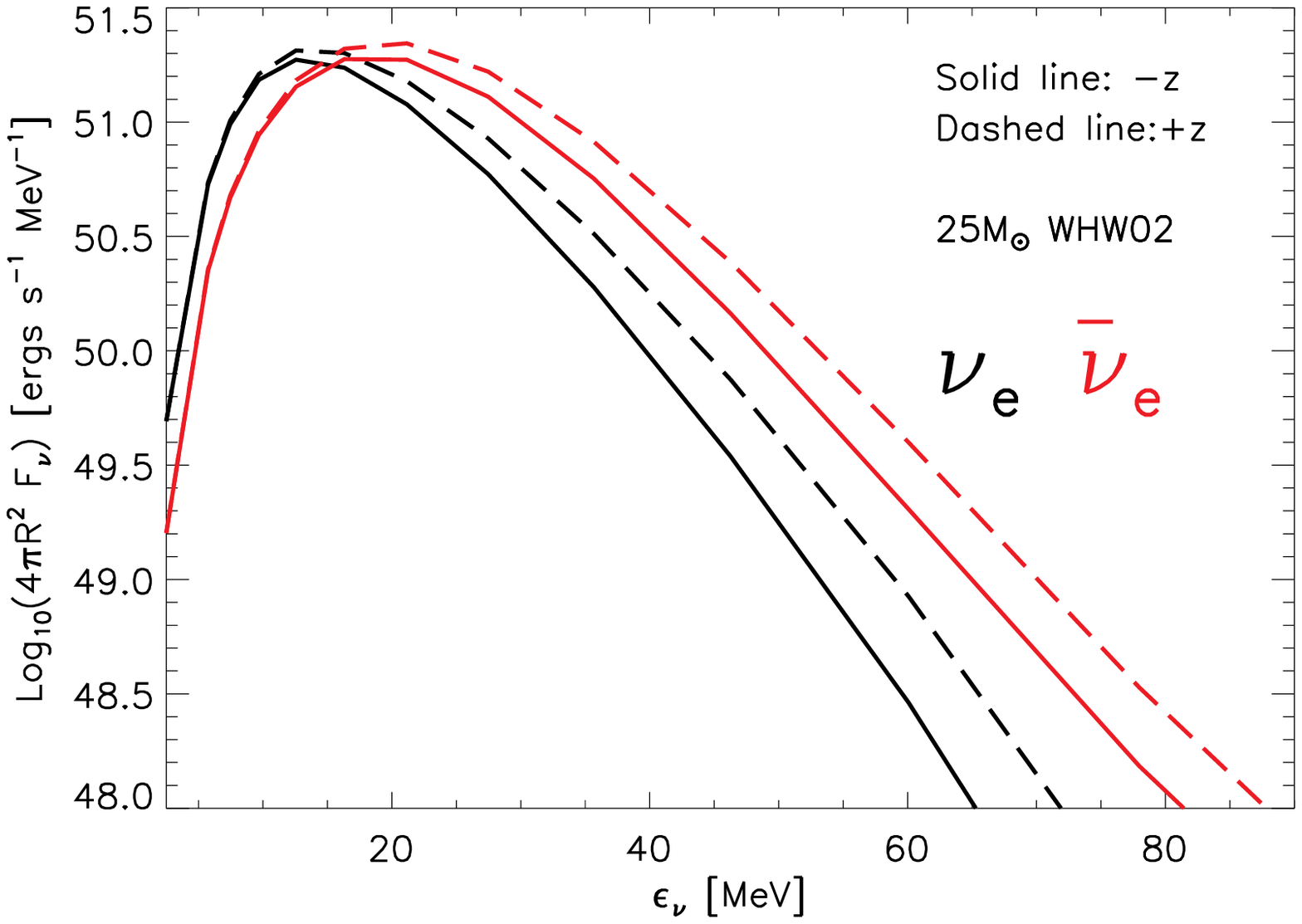}
\caption{{\it Left:} Energy spectra of the electron-neutrino (black) and
anti-electron neutrino (red) fluxes along the poles (solid line: negative
$z$-direction; dashed line: positive $z$-direction) for the 13-M$_{\odot}$
model of WHW02.  The fluxes are multiplied by a factor $4 \pi R^2$ and 
are at 1.44\,s after bounce. {\it Right:} Same as at the left, but for
the 25-M$_{\odot}$ model of WHW02 at 1.42\,s after bounce. Note that the hotter and
higher fluxes are in each case in the direction of the explosion, though the two 
models shown explode in different directions (see Fig. \ref{fig:5}).}
\label{fig:11}
\end{figure*}

With VULCAN/2D we are also able to ascertain for the first time the
magnitude and sign of the impulse due to anisotropic neutrino emissions. We find that
during our simulations (to approximately 1.5 seconds after bounce)
the neutrino recoil effect on the core is not large, at most $\sim$50 km s$^{-1}$, but that by
the end of our simulations it is still growing and is in the opposite direction
to the blast.  Hence, after the explosion commences, the impulses on the
protoneutron star due to the matter ejecta and the neutrino radiation {\it add}. Figure
\ref{fig:10} depicts both the net force and the accumulated impulse due to neutrinos
during our simulation of the post-bounce phase of the 13-M$_{\odot}$ model of WHW02.
The negative sign indicates that the neutrinos are emerging preferentially
in the direction of the exploding matter (in this case, ``downward"; 
see Fig. \ref{fig:5}), and not towards the accreting side.
The small magnitude of the neutrino force during the delay to
explosion may seem inconsistent with the very anisotropic accretion.
However, the radiation field is much smoother by its nature than the material
field. Importantly, the neutrinos are not radiated instantaneously upon compression in an
accretion column onto the protoneutron star.  The matter is too opaque for immediate
reradiation.  Rather, the neutrinos emerge after the compressed accreta have
spread more uniformly over the inner core, and, therefore, are radiated much more isotropically
than the matter is accreted.  After explosion, the neutrinos can emerge more easily
along the direction of the blast, since the material around the neutrinospheres
thins out in this direction, and not along the direction experiencing continuing accretion,
which, as stated, is more opaque.  This result is at odds with the conclusion
of Fryer (2004).  Figure \ref{fig:11} depicts the spectra of the 
$\nu_e$ and $\bar{\nu}_e$ neutrinos in the up and down
directions (along the poles) near the end of the simulations of 
the 13-M$_{\odot}$ and 25-M$_{\odot}$ models.  As Fig. \ref{fig:11} demonstrates,
the radiation is {\it hotter} in the direction of the blasts, which for these
models are in opposite directions (see Fig. \ref{fig:5}).  It is also ``brighter"
in those directions.  This is consistent with the sign of the neutrino recoils shown  
in Fig. \ref{fig:10} and our statement that neutrino impulses and wind recoils 
add. Hence, we conclude that asymmetric neutrino recoil, integrated even longer than
we have in this study, can contribute significantly and naturally 
to the final pulsar kick.  However, we can not, at this stage, determine 
whether the matter recoil or the neutrino recoil will eventually prove the more important.  
Nevertheless, what has emerged from our simulations is a straightforward mechanism 
for imparting a sizable kick to the residue protoneutron star that does not
require anything but the asymmetric explosion that arises naturally in 
our calculations without exotic physics.

\section{Discussion and Conclusions}
\label{discussion}

The acoustic and core oscillation mechanism we study in this paper
and in Burrows et al. (2006) has a number of features that 
distinguish it from the classic neutrino mechanism.
Here, we provide a list of some of its more salient aspects. 
In arriving at this list, 
we have been guided by the long-term ($\sim$1.5 seconds) 
simulations we have performed for this paper that use the 
MGFLD version of VULCAN/2D (see Appendix), liberate the core, and
include the full 180$^{\circ}$ angular domain.  Calculations that do not have the latter two features,
that do not include neutrino transfer, or that do not go to very late times can not be used
to study the core-oscillation/acoustic component.  
We find that:

\begin{itemize}

\item The SASI, when neutrino losses are properly included, does not lead to explosion
in and of itself, but creates an anisotropic accretion regime onto the core that
eventually leads to nonlinear core pulsation with symmetries ($\ell$s) similar to those of the
most unstable SASI modes themselves. Both $\ell = 1$ and $\ell = 2$ core 
g-mode oscillations can be the most prominent, though the $\ell = 1$ 
mode arises earlier and generally dominates (Burrows et al. 2006).

\item The acoustic power generated by the core oscillation seems to be dumped 
into the mantle for many seconds after bounce, longer than the standard 
neutrino mechanism is thought to operate.  Furthermore, it can take many
hundreds of milliseconds to $\sim$0.6 seconds (determined by the progenitor mass
accretion rate) before the core oscillation itself achieves large amplitudes. 

\item During the early phase of the explosion, the acoustic power steadily punches
out through the accreta and generates a collimated explosive flow.  This is aided
by neutrino heating, which if the acoustic component were suppressed
would be the standard underenergetic neutrino-driven wind mechanism (Burrows \& Goshy 1993;
Buras et al. 2006ab).  The early net velocities of this flow are low,
but before and during this phase a cavity is filled with acoustic energy radiated by the core oscillation.
Sound waves bounce off the cavity walls and reverberate in the cavity.

\item At later times, during what would have been the neutrino-driven
wind phase in the traditional neutrino-driven explosion, the acoustic mechanism is still aided by
neutrino heating at the level of $\sim$10$^{50}$ ergs.

\item The inner, early blast is mostly unidirectional and is naturally collimated by the accretion flow
that is parted by the blast and diverted to the opposite side of the protoneutron star.  There is simultaneous
accretion, explosion, and core oscillation (Burrows et al. 2006).

\item The matter that punches out in the explosion experiences a Kelvin-Helmholtz shear
instability which rolls up the interface between the ejecta and the cocooning accreta as the wide-angle
``jet" (the early explosion) emerges. However, the coarse outer zoning of our calculations
at large radii (3000-6500 kilometers) is currently insufficient to resolve this
interesting phenomenon properly.

\item The explosion is very radiation-dominated; most of the explosion energy is initially
in internal energy, not kinetic energy.  Furthermore, the ejecta have high entropies
(100-1000 per baryon per Boltzmann's constant, generated by
both neutrino heating and acoustic power), far larger than the generic values (10-50)
associated with the early-phase neutrino mechanism.  Hence, if the acoustic mechanism works,
the early development of the explosion into the star and the associated explosive nucleosynthesis
can not properly be simulated with a piston or a ``kinetic-energy" bomb.

\item The high entropies suggest that some of the ejecta will undergo
r-processing (Hoffman et al. 1996; Woosley et al. 1994).  This is
the first time numerical supernova explosions have simultaneously
and naturally generated the conditions that may be necessary for the r-process.

\item Our calculations are Newtonian and have been 
done only with the equation of state of Shen et al. (1998).
General relativity will change the core oscillation frequencies,
and so will affect the acoustic power, its evolution, and the timing of the various phases.
Consequently, it should eventually be included in the simulations.  The incompressibility of nuclear
matter will also affect the g-mode frequencies and, hence, a study of the dependence
on the nuclear equation of state would be illuminating and may provide diagnostics
of the EOS at high densities.

\end{itemize}

When the SASI is in its vigorous non-linear phase, the $\ell = 1$ oscillations
result in quasi-periodic fluctuations in the effective accretion rate and ram pressure on any given
side of the inner core.  In the canonical neutrino-driven mechanism of supernova
explosions, when and after the explosion occurs the pressure around the neutrinospheres
decays.  When this pressure is sufficiently low, a neutrino-driven wind spontaneously
emerges from the inner core, announced and preceded by a secondary shock wave (Burrows 
\& Goshy 1993; Burrows, Hayes, \& Fryxell 1995).  This is what happens in the standard
neutrino-driven scenario when the flow is semi-spherical. However, the SASI 
can set up a situation in which the pressure and ram pressure on one side
are such that that side of the core becomes unstable to the emergence of 
a neutrino-driven wind even before the canonical explosion. In fact, this wind can 
be the explosion itself and need not be preceded 
by a primary explosion.  This is what Buras et al. (2006b) see for their 11.2-M$_{\odot}$
simulation.  However, such an explosion seems generically underenergetic. 
In the acoustic mechanism, the neutrinos are replaced/dominated by the acoustic power,
but the general paradigm in which the SASI leads temporarily/periodically to 
lower pressures on one side of the core that enable the emergence of an asymmetric wind
still obtains.  In any case, an aspherical ``wind" is a good description 
of the supernova explosion (see Scheck et al. 2004,2006 
and Burrows \& Goshy 1993) and the breaking of spherical symmetry
is the key.  The latter can also enable simultaneous accretion and explosion,
thereby solving the problem of the accretion tamp that has bedeviled 
the theory of the neutrino mechanism for years.

We see in the breaking of spherical symmetry in our simulations
and in the unipolar nature of the resulting explosions a natural explanation for the 
polarizations observed in the inner debris of Type Ic (Wang et al. 
2003) and Type II (Leonard et al. 2006) supernovae.
Inner ejecta asymmetries of 2:1 or 3:1 are easily obtained in this model,
and in fact in all modern non-MHD explosion models, and do not require MHD
jets.

What the actual and relative contributions of sound and neutrinos are 
to the supernova phenomenon as a function of progenitor 
remains to be determined and will require even more sophisticated
numerical tools than we have applied here to reach a definitive answer. 
The calculations presented in this paper have several limitations (see Appendix).
We are doing them in 2D; 3D, while out of reach in the short term,
will be necessary in the long term.  We have employed a MGFLD, not a multi-angle,
formulation, and the Doppler-shift terms in the transport equation
have been dropped.  While these velocity-dependent terms are very different in the 
laboratory-frame formulation (Hubeny \& Burrows 2006) 
we have adopted than in the comoving frame formulation of 
Buras et al. (2006a), they should nevertheless be incorporated. 
We have used only 16 energy groups; using more ($\sgreat 20$) is preferred (Thompson, Burrows, \& Pinto 2003).
The spatial resolution in the center is good, but can be improved on the outside
exterior to $\sim$200 km. The calculations and gravitational field are Newtonian;
we expect that general and special relativistic effects can be important
on the inside and outside, respectively. The opacities employed 
are sophisticated, but the neutrino-matter correlation effects
at higher densities need a second look.  

All in all, the supernova problem
has resisted attempts at resolution for too long.  In the calculations
we present here, and in Burrows et al. (2006), we see new, perhaps provocative, 
ideas emerge that will require fresh approaches to test and verify them.     
A potentially new role for the inner core has been highlighted and
the intriguing suggestion that acoustic power might compete with neutrino power
to ignite the supernova explosion has been put forward.  Furthermore, 
we find that there is much to explore in the interaction of the core and shock
instabilities.  Whether these new ingredients in supernova theory  
are keys, or curiosities, awaits the next generation of simulations.

\acknowledgments

We thank Todd Thompson, Rolf Walder, Stan Woosley,  
John Blondin, and Thierry Foglizzo for fruitful discussions and 
their insight. We acknowledge support for this work
from the Scientific Discovery through Advanced Computing 
(SciDAC) program of the DOE, grant number DE-FC02-01ER41184,
and from the NSF under grant AST-0504947.
E.L. thanks the Israel Science Foundation
for support under grant \# 805/04, and C.D.O. thanks the LSU Center for Computation
and Technology for providing CPU time on their SuperMike Linux cluster.
J.W.M. thanks the Joint Institute for
Nuclear Astrophysics (JINA) under NSF grant PHY0216783 for providing a 
graduate fellowship. This research used resources of the National 
Energy Research Scientific Computing Center, which is supported by the
Office of Science of the U.S. Department of Energy under Contract No. DE-AC03-76SF00098.
We are happy to acknowledge the National Center for Computational Sciences
at Oak Ridge for a generous allocation of computer time on Jaguar.
Finally, we thank Don Fisher, Youssif Alnashif, and Moath Jarrah
for their help generating both color stills and movies associated with this work
and Jeff Fookson and Neal Lauver of the Steward Computer Support Group
for their invaluable help with the local Beowulf cluster Grendel.

\begin{appendix}

\section{VULCAN/2D: A Multi-Group, Multi-Angle Rad/Hydro Code and its MGFLD Variant}
\label{vulcan2d}

In this appendix, we assemble paragraphs on some of the numerical techniques used in VULCAN/2D,
in particular its Multi-Group, Flux-Limited Diffusion realization.  Some of this discussion can be found
in our other papers using VULCAN/2D (e.g., Livne et al. 2004; 
Ott et al. 2004; Burrows et al. 2006; Walder et al. 2005;
Dessart et al. 2006ab).  We believe that assembling this technical information in one
place will better help the reader understand the computational issues that
surround such supernova codes in general, and VULCAN/2D in particular.
Importantly, though in the past workers focussed on improvements
in neutrino transfer and transport, the acoustic mechanism requires special attention
be paid to the hydrodynamics, grid structure, momentum conservation, gravity solvers, and, we suggest,
a moving grid as well, to ensure and maintain good resolution in the inner core.  VULCAN/2D
is the first supernova code for which these issues have been central considerations.

The code VULCAN/2D uses the explicit hydrodynamic approach described in Livne (1993),
with the implicit transport methods discussed in Livne et al. (2004) and Walder et al. (2005).
It is a Newtonian, 2D, multi-group, multi-angle radiation/hydrodynamics
code\footnote{Hence, it is a 6-dimensional (1(time) + 2(space) + 2(angles) + 1(energy groups)) solver.}
with an Arbitrary-Lagrangean-Eulerian (ALE) structure (with remap),
a scalar von Neumann-Richtmyer artificial viscosity scheme to handle shocks, and a
fast Multi-Group, Flux-Limited Diffusion (MGFLD) variant.  The full Boltzmann version
discretizes the angular variables using the discrete-ordinates (S$_n$) method.
The code can handle axisymmetric rotation. Velocity terms in the transport sector, such as
Doppler shifts, are not included in the code, though advection is.  Note that the velocity
terms in Eulerian transport are different from the corresponding terms in the comoving frame
and that general statements about their relative importance are very frame-dependent.
We parallelize only in energy groups using MPI and in 2D no domain decomposition is required.
As a result, and in practice, VULCAN/2D is very scalable and the communication overhead
is only 2\% to 8\% of the total run time.  The fact that domain decomposition
(such as is used in FLASH and CACTUS) is not necessary,
that we can achieve almost perfect parallelism in energy groups, and that we can
include rotation, has enabled us to achieve a viable 2D simulation
capability.

Note that energy redistribution due to inelastic electron scattering is of only modest 
import on infall, affecting the trapped electron fraction (Y$_e$) and entropy ($S$) by only $\sim$$10$\%.
Furthermore, at a neutrino energy of 10 MeV, the neutrino-electron scattering cross section 
is $\sim$100 times smaller than the dominant cross sections off nucleons.  Hence, we have not felt 
it urgent to include into VULCAN/2D energy redistribution by neutrino-electron scattering.
Fortunately, such energy redistribution, because it is sub-dominant, can be
handled semi-explicitly (Thompson, Burrows, \& Pinto 2003), thus avoiding
interprocessor communication during an implicit solve. A scheme for this is
already written and debugged, since it is used in SESAME
(Burrows et al. 2000; Thompson, Burrows, \& Pinto 2003), and is quite stable.
The attempts by others to handle the full energy/angle redistribution
problem implicitly have resulted in codes that are thereby slower by many {\it factors} (not percent),
severely inhibiting their use for explorations in supernova theory.

In 2D, the calculations are axially/azimuthally symmetric,
and we use cylindrical coordinates ($r$ and $z$), but the grid points
themselves can be placed at arbitrary positions.  This allows us to employ
a Cartesian grid at the center (typically, the inner $\sim$20-30 kilometers) and transition
to a spherical grid further out.  The grid resolution is essentially
uniform everywhere within this $\sim$20-30\,km.  A version of this grid structure is plotted
in Ott et al. (2004).  The Cartesian format in the interior allows us to avoid
the severe Courant problems encountered in 2D by other
groups employing grid-based codes due to the inner angular
Courant limit, to enable core translational motion, and, thereby, to perform the calculations
in full 2D all the way to the center.  In many simulations to date, the inner core
has been calculated in 1D and grafted onto an outer region that was handled in 2D (e.g., Burrows, Hayes,
\& Fryxell 1995; Janka \& M\"{u}ller 1996; Buras et al. 2003,2006ab;
Swesty \& Myra 2005ab) or has been excised completely (e.g., Blondin, Mezzacappa, \& DeMarino 2003;
Blondin \& Mezzacappa 2006; Scheck et al. 2004,2006).  The gray 
SPH simulations of Herant et al. (1994) and Fryer \& Warren (2002,2004)
are an exception.  Originally, a major motivation for this global 2D
feature was the self-consistent investigation of core translational motion and
neutron star kicks.  However, freeing the core has the advantage that no
other multi-group supernova code has of simulating the oscillation of the core
and its acoustic radiation.  

Note that due to the grid singularity in spherical coordinates at $r=0$
and the inherent difficulties of constructing a reliable finite-difference scheme
and boundary conditions at that singularity (reflecting?), 
codes that attempt to include the core in 2D or 3D using
spherical coordinates are likely to artificially inhibit 
translational motion there and, thus, to inhibit $\ell = 1$ g-modes.   Even a simple
Galilean transformation/translation of a hydrostatic core, which is what 
our special grid was designed for, may not be easy when using the standard realizations
of a spherical grid in 2D/3D.

Outside the Cartesian mesh, our baseline calculations have 
typically employed 121-180 angular zones equally spaced over the 
entire 180$^{\circ}$ of the symmetry domain, and $\sim$160 radial shells 
logarithmically allocated between $\sim$20\,km (generally 10, 
20, or 30 km) (Dessart et al. 2006b) and the outer radius at 6400$-$10000\,km.
Along the symmetry axis ($r = 0$), we use a reflecting boundary, while at the outer
boundary, we use either an outflow or a $v=0$ boundary condition for the matter and a
free-streaming boundary condition for the neutrinos.

\subsection{2D Multi-Group Flux-Limited Diffusion of Neutrinos}
\label{mgfld}

The MGFLD implementation of VULCAN/2D is fast and uses a vector
version of the flux limiter found in Bruenn (1985) (see also Walder et al. 2005).
Using the MGFLD variant of VULCAN/2D allows us to perform an extensive
study that encompasses the long-term evolution of many models. However, 
MGFLD is only an approximation to full Boltzmann
transport and differences with the more exact treatment will emerge
in the neutrino semi-transparent and transparent regimes above the 
protoneutron star surface. Nevertheless, inside the neutrinospheres the
two-dimensional MGFLD approach provides a very reasonable
representation of the multi-species, multi-group neutrino radiation fields.

The evolution of the radiation field is described in the diffusion approximation
by a single (group-dependent) equation for the average intensity $J_g$ of energy group $g$ 
with neutrino energy $\varepsilon^g_{\nu}$:
\begin{equation}
  \dJgdt - div(D_g \nabla J_g) + \sigma^a_g J_g = S_g \, ,
\label{diff}
\end{equation}
where the diffusion coefficient is given by $D_g={1 \over 3\sigma_g}$ (and then is flux-limited
according to the recipe below), the total inverse mean-free-path 
(``cross section") is $\sigma_g$, and the inverse absorption 
mean-free-path (absorption ``cross section") is $\sigma^a_g$. 
The source term on the R.H.S. of eq. (\ref{diff}) is the
emission rate of neutrinos of group $g$.
Note that eq. (\ref{diff}) neglects inelastic scattering between
energy groups. 

The finite difference approximation for eq. (\ref{diff}) consists of cell-centered
discretization of $J_g$. It is important to use cell-centered discretization
because the radiation field is strongly coupled to matter and the thermodynamic
matter variables are cell-centered in the hydrodynamical scheme. The finite-difference 
approximation of eq. (\ref{diff}) is obtained by integrating the equation over
a cell. Omitting group indices and cell indices one gets:
\begin{equation}
  V[\cdt (J^{n+1}-J^n) + \sigma^a J^{n+1}] +{\Sigma \dsi \cdot \BF_i^{n+1}} = V S \, .
\label{diffmgfld}
\end{equation}
Here $V$ is the volume of the cell, $\dsi$ is the face-centered vector ``$area_i \bf{n_i}$,"
$\bf{n_i}$ being the outer normal to face $i$. The fluxes $\BF_i$ at internal faces
are the face-centered discretization of
\begin{equation}
 {\BF_i} = - D_i \nabla J^{n+1} \, ,
\label{flux}
\end{equation} 
where
\begin{equation}
 D_i = FL [{1 \over 3\sigma_i}]  \, .
\end{equation}
Our standard flux limiter, following Bruenn (1985) and Walder et al. (2005), is
\begin{equation}
 FL [D] = {D \over {1 + D |\nabla J| /J}}
\end{equation}
and approaches free streaming when $D$ exceeds the intensity scale height
$J/|\nabla J|$. The fluxes on the outer boundary of the system are defined
by free streaming outflow and not by the gradient of $J$.
Note that in eq. (\ref{flux}) the fluxes are defined as face quantities, so that
they have exactly the same value for the two cells on both sides of that face.
The resulting scheme is, therefore, conservative by construction. In order to have
a stable scheme in the semi-transparent regions (large $D_g$) we center the variables
in eq. (\ref{diffmgfld}) implicitly. The fluxes, defined by the intensity at the end
of the time step, couple adjacent cells and the final result is a set of 
linear equations. The matrix of this system has the standard band structure
and we use direct LU decomposition to solve the linear system. For a moderate grid
size the solution of a single linear system of that size does not overload
the CPU.

We have parallelized the code according
to energy groups. Each processor computes one to a few groups (usually one) and
transfers the needed information to the other processors using standard MPI
routines. In our baseline runs, we employ 16 energy groups per neutrino 
species, logarithmically spaced from 1 or 2.5 MeV to 250 or 320 MeV.
Since we do not split the grid between processors, the parallelization 
here is very simple. In fact, each processor performs the hydro step on
the entire grid. In order to avoid divergent evolution between different 
processors due to accumulation of machine round-off errors, we copy the 
grid variables of one chosen processor (processor 0) into
those of the other processors typically every thousand steps.

\subsubsection{Coupling Radiation to Matter}
\label{coupling}

The numerical scheme used in VULCAN consists of a Lagrangean step,
followed by a remapping step to the Eulerian grid (Livne 1993).  This makes the
code similar in this regard to traditional ALE (Arbitrary-Lagrangean-Eulerian)
codes.  The hydrodynamical variables are all
cell-centered, except for the position and the velocity, which are node-centered.
The variables of the radiation field are also cell-centered, so that the
interaction between the radiation field and matter is properly centered.

The time advancement in both the radiative and the
hydrodynamical sectors is computed in the Lagrangean step,
whereas the remapping step changes only the spatial discretization of the variables
over the numerical grid. We describe here only the
Lagrangean scheme.  The transport equation itself, for a
given source, is always computed in a fully implicit manner.

We first advance the velocity by half a timestep:
\begin{equation}
\vel^{n+1/2}=\vel^{n}+0.5\dt(-\frac{\nabla p^n}{\rho^n} -\nabla U_G) \, ,
\label{18e}
\end{equation}
where $p$ is the pressure, $\rho$ is the mass density, $\Delta t$ is the timestep, and 
$U_G$ is the gravitational potential (but see the subsection on Gravity and Poisson Solvers below).
The position vector is then advanced using
\begin{equation}
\pos^{n+1}=\pos^{n}+\dt\vel^{n+1/2}  \, .
\label{19e}
\end{equation}
Denoting by $V$ the volume of a cell, Lagrangean mass conservation takes the form:
\begin{equation}
\rho^{n+1}=\rho^n {V^n \over V^{n+1}}     \, .
\label{20e}
\end{equation}
We then solve the adiabatic energy equation for the specific internal energy:
\begin{equation}
e^*=e^n-1/2(p^*+p^n)({1 \over \rho^{n+1}}-{1 \over \rho^{n}}) \, .
\label{21e}
\end{equation}
Equation (\ref{21e}) is iterated to convergence.
At this stage, we compute new cross sections and emission sources
\begin{equation}
S_{em}=\sum_g{\sigma^a_g J_g^{eq}}    \, ,
\label{22e}
\end{equation}
where $J_g^{eq}$ is the LTE intensity (a function of density,
temperature, and composition).
Using those cross sections and sources, we solve the transport equation. 
The net change in the radiation energy density is given by:
\begin{equation}
\Delta E_r=\dt \sum_g{\sigma^a_g (J_g^{eq}-J^{n+1}_g)}
\label{23e}
\end{equation}
and this is also minus the net change in the matter energy density. Consequently, we compute
the final energy density, pressure, and temperature using
\begin{equation}
e^{n+1}=e^*-\Delta E_r/\rho^{n+1}
\label{24e}
\end{equation}
and the equation of state.

For the supernova problem, we also need to compute the degree of neutronization of matter
due to electron capture and other charged-current processes (Burrows \& Thompson 2004).
We obtain
\begin{equation}
Y_e^{n+1}=Y_e^n - \dt [\sum_g{\sigma^a_g (J_g^{eq}-J^{n+1}_g) /\varepsilon^g_\nu}] {1 \over N_a \rho} \, ,
\label{25e}
\end{equation}
where $N_a$ is Avogadro's number and $Y_e$ is the electron fraction.
Finally, we advance the velocity due to the new matter pressure by a further half timestep
and due to the radiation pressure ($\mathbf F_{rad}^{node}$) by a full timestep:
\begin{equation}
\vel^{n+1}=\vel^{n+1/2}+0.5\dt [{1\over \rho^{n+1}}(-\nabla p^{n+1}+2\mathbf F_{rad}^{node})
   -\nabla U_G] \, .
\label{26e}
\end{equation}
The radiation force at grid nodes is evaluated by a simple averaging process
using the radiation force at cell centers and the definition of the radiation flux (eq. \ref{flux}).

\subsection{Advection of Angular Momentum}
\label{angular}

Since VULCAN/2D uses an Eulerian grid, when we study rotating models the specific angular momentum
is advected with the flow in the same manner as linear momentum
components. In so doing, we maintain global angular momentum
conservation by construction.  Note that the axis in cylindrical
coordinates is a singularity and, as such, is prone to slightly larger
errors than can be expected elsewhere on the grid. However, the actual
volume of the cells nearest the axis is very small and the
errors do not affect the overall flow. In the past
(Walder et al. 2005) we have estimated such departures near the singularity
for rotating models to be no more than $\sim$10\% in any flow variable,
and to be much smaller elsewhere.

\subsection{Gravity and Poisson Solvers}
\label{poisson}

Gravity is a key force in multi-dimensional astrophysical hydrodynamics.
However, many calculations in the past have employed only
the monopole term and/or have complemented the gravitational
force term ($F_G$), written as a gradient of a potential
in the momentum equation (eq. \ref{18e}) with a corresponding $\vec{v}\cdot F_G$
term in the energy equation.  The latter approach is perfectly reasonable, but
given the inherently approximate
nature of finite-difference realizations of the partial differential equations, does not
guarantee momentum conservation, nor consistency between the momentum and energy
equations when written in Eulerian form.
To address this, we have implemented a version of the code in which the $z$-component of gravity appears
as the divergence of a stress tensor (Shu 1992; Xulu 2003).  This ensures, in principle,
the conservation of momentum in that direction, or at least guarantees that
in fact the gravitational force of mass parcel ``A" on ``B" is equal and opposite
to the gravitational force of mass parcel ``B" on ``A."  

\begin{figure*}
\figurenum{12}
\plotone{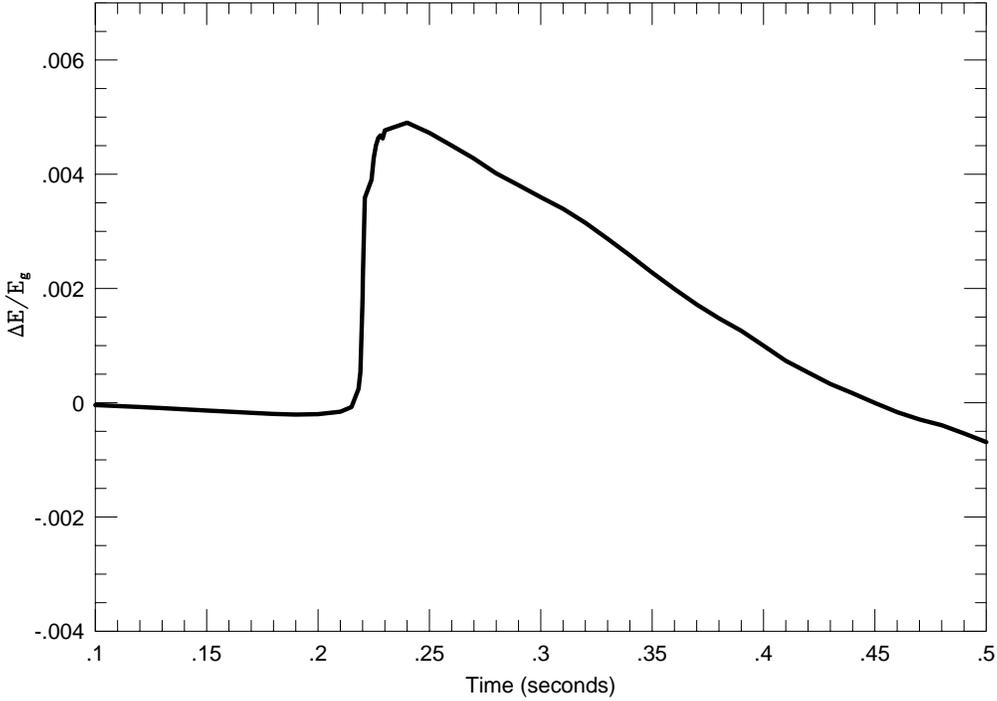}
\caption{A plot of $\Delta E/E_g$
versus time (from the start of the calculation),
where $\Delta E$ is the total energy conservation error and
$E_g$ is the gravitational potential energy. This figure is 
for the simulation performed by Ott et al. (2006a) of the 
11-M$_{\odot}$ progenitor model of WW95 that was rotated
to have a rapid initial spin of 2.68 rad s$^{-1}$ in the core.
For rotating models, energy conservation will generally be 
worse than for non-rotating models. The dimensionless
ratio $\Delta E/E_g$ is a useful measure of the degree to which energy
is conserved during a simulation using VULCAN/2D.  The
bump on the plot at $\sim$0.22 seconds occurs at bounce. See text in the Appendix 
under ``{\it Gravity and Poisson Solvers}" 
for a discussion.
}
\label{fig.append}
\end{figure*}

Currently, there are in VULCAN/2D two Poisson solvers - a \it multipole solver \rm and 
a \it grid solver \rm . The multipole solver is a standard Legendre expansion
and we typically employ 20 to 33 terms. For the potential calculations, 
one generally needs a special auxiliary grid, which is 
not identical with that of our complex hydro grid. This leads to a number of interpolations 
between the grids, which can introduce significant numerical errors. Most 
importantly, with the multipole solver conservation of total energy is poor through bounce and later.
Conservation of total energy is much better with the other solver, the grid solver, and the numerical noise
in the core region is significantly reduced when we employ it. However, due to
the unavoidable operator split between the hydro and the 
gravity calculations it is generically hard to get good total energy conservation.
VULCAN/2D generally conserves energy to an average of better than
0.4\% in terms of $\Delta$E/E${_\mathrm{grav}}$, with the worst 
energy conservation phase near core bounce.  
Figure \ref{fig.append} depicts $\Delta$E/E${_\mathrm{grav}}$
  versus time for the published 11.0-M$_{\odot}$ run from Ott et al.
  (2006a) that includes neutrinos and has a rapid initial spin of
  2.68 rad s$^{-1}$ in the core. Rotation generally increases the
  error. The major reasons $\Delta$E/E${_\mathrm{grav}}$ is not zero
  are: 1) the code is not automatically conservative and, hence,
  the gravitational term in the energy equation the code ``thinks"
  it uses, given the finite difference approach, is different from
  our post-processed calculation of $\int \frac{1}{2}\rho\Phi {\rm dV}$.
  Differences of a percent in this can cause large differences that
  may or may not be meaningful; 2) A similar point can be made
  concerning the neutrino energy integration for logarithmically
  distributed energy gridding: How accurately can one integrate under
  a curve that is sparce at higher energies? 3) The 2D ALE code uses
  a remap step that is not ``perfect" when the velocities change fast
  near bounce; 4) A predictor/corrector step, which we don't have, would
  give us higher order accuracy in time; and 5) There are slight differences
  in the finite-difference treatment of the matter and radiation source
  terms, which are formally equal and opposite.  For a comparison,
  Liebendorfer et al. (2004) quote errors of $\sim$0.005 for the same
  quantity at the end of their 1D calculations, similar to the peak
  problem we show in Fig. \ref{fig.append}.

The grid solver, which we employ in this paper, uses the standard finite-element method (FEM), which is adequate 
for unstructured grids, to get the potential at grid nodes.
In axial symmetry Poisson's equation takes the form
\begin{equation}
 \Delta \Phi = \ovr \frac{\partial}{\partial r}(r\frac{\partial\Phi}{\partial r}) + \frac{\partial^2 \Phi}{\partial z^2}=-4
\pi G \rho \, .
\label{new1}
\end{equation}
Let $\{\alpha_i(r,z)\}$ be a set of interpolation functions which span our FEM approximation.
Multiplying eq. (\ref{new1}) by $\alpha_i$ and integrating by parts over the entire domain one gets :
\begin{eqnarray}
 - \int\int \nabla\Phi\nabla\alpha_i r dr dz + \{surface-integral\} 
\nonumber\\
 = -4\pi G \int\int \rho \alpha_i r dr dz \, .
\end{eqnarray}
In particular, if we expand $\Phi$ using the set $\{\alpha_i\}$, specifically
\begin{equation}
 \Phi(r,z)= \Sigma_j \Phi_j\alpha_j(r,z) \, ,
\end{equation} where $\Phi_j$ is the value of $\Phi$ at node j, we get  a linear system of the form
\begin{equation}
 A{\bf\Phi} = \bf B \, ,
\end{equation} where
\begin{equation}
 A_{ij} = - \int\int \nabla\alpha_i\nabla\alpha_j r dr dz
\label{new2}
\end{equation} and
\begin{equation}
 B_i = -4\pi G \int\int \rho \alpha_i r dr dz  \, .
\label{new3}
\end{equation}
The matrix $A$ has good qualities if we choose $\{\alpha_i\}$ to be continuous, positive,
 and local, with the following specifications:
 $ \alpha_i(r_j,z_j) = \delta_{ij}$ at the nodes of the grid, $\alpha_i = 0$ in any zone not
 containing node $i$, and $\Sigma_i \alpha_i(r,z)=1$ everywhere.
 In practice, we employ bilinear interpolation functions in each zone.
 The integrals (\ref{new2}) are computed once, using numerical integration, and the integrals (\ref{new3})
are computed each timestep.

Note, however, that the grid solver needs boundary values, and for this we take the
zero'th moment $\Phi_b=-G M/R_b$, where $R_b$ here is the distance between a boundary point
and the center of mass (which is usually very close to the center of the grid). This
approximation is good for large outer radii where the potential drops by orders of magnitude
compared with that at the center.

\subsubsection{Gravity and the Conservation of Linear Momentum}
\label{linear}

Importantly, we incorporate the
gravitational force along the symmetry axis in an automatically
momentum-conserving fashion by writing it in divergence form.  

Gravity enters the momentum equation
\begin{equation}
 { \partial(\rho\mathbf{v}) \over {\partial t}} = -\nabla P + \rho \nabla \Phi
\label{1}
\end{equation}
via the potential $\Phi$ defined by the Poisson equation
\begin{equation}
  \Delta \Phi = -4\pi G \rho \, .
\label{poissoneq}
\end{equation}
In eq. (\ref{18e}), the gravity term is not in conservative form and, therefore, it is not
useful for obtaining a conservative numerical scheme. To overcome this, we
use eq. (\ref{poissoneq}) to obtain
\begin{equation}
 {\bf{f_g}} = \rho  \nabla \Phi = -\Gpi \Delta \Phi \nabla \Phi \, .
\label{pforce}
\end{equation}
In Cartesian coordinates, the R.H.S. of eq. (\ref{pforce}) can be written in a divergence form:
\begin{equation}
  \bf{f_g} = div (S) \, ,
\label{divs}
\end{equation} 
where the \it gravitational stress tensor \rm $\mathbf S$ is derived from $\Bg = \nabla \Phi $
(Shu 1992, p. 47):
\begin{equation}
  {\mathbf S_{ij}}  = -\Gpi (g_i g_j - \half \mid\Bg\mid^2 \delta_{ij}) \, .
\label{sij}
\end{equation}

In cylindrical coordinates (r, z) with axial symmetry we use the same idea with the
Laplacian $\Delta \Phi= {1\over r}{\partial(r g_r)\over \partial r} + {\partial(g_z)\over \partial z}$.
Direct calculation yields:
\begin{equation}
   f_z = -\Gpi \Bigl({1 \over r} {\partial(r g_r g_z) \over \partial r} + \half {\partial(g_z^2-g_r^2) \over \partial z}\Bigr)
\end{equation}

and

\begin{equation}
   f_r = -\Gpi \Bigl( {1 \over 2r} {\partial[r(g_r^2- g_z^2)] \over \partial r} + {\partial(g_r g_z) \over \partial z}+{1 \over 2r} (g_r^2+g_z^2) \Bigr) \, .
\end{equation}

Note that the z-component of the momentum equation (eq. \ref{18e}) now has a 
divergence form and, therefore, can be integrated to give
a conservative finite-difference scheme.  

In practice, we compute the potential on grid nodes and then
compute $\bf g$ at cell centers. The forces are computed by integrating eqs. (\ref{divs}) and (\ref{sij})
over a control volume around a node, where the boundary line of this control volume
passes through the centers of the cells circling that node. The conservative form of
$f_z$ expresses itself in the scheme by having contributions from boundary terms only.

\subsection{Grid Motion}
\label{gridmotion}

In full 180$^{\circ}$ simulations the core has the freedom to escape the center of the grid,
where the resolution is finest. Experience shows that in very long simulations the
motion of the core off the center of the grid is numerically unstable and can lead
to an artificial ``kick." In order to avoid this situation, VULCAN/2D 
has an option to move the grid after bounce to maintain the best zoning under the core, whether
it moves or not, while at the same time tracking this core motion.  This feature
ensures that the highest resolution is placed under most of the mass. Adding 
a constant $\delta v_z$ to $v_z$ everywhere does not change anything in the dynamics.
Generally, we calculate the position of the center of mass of the inner material above
a density of 10$^{12}$ g cm$^{-3}$ and execute the grid motion every timestep
to position the grid center at this point.  
As a test, we have allowed grid motion every 100 timesteps and
the results fall right on top of those done every timestep.  This procedure typically
ensures that the center-of-mass stays at the center of the grid to within $\sim$10-100 {\it meters}.

\subsection{Seeds for Instabilities}
\label{seed}

The instabilities that develop in the early stages of the post-bounce
phase are seeded by the slight perturbations introduced due to the
non-orthogonal shape of the grid regions that effect the transition
from the inner Cartesian grid to the outer spherical grid (see Fig.
4 in Ott et al. 2004) and by noise at the one part in $\sim$10$^6$ level
in the EOS table interpolation.  Since the resulting turbules execute
more than twenty overturns during the initial phase of convective
instability, and this convective phase reaches a quasi-steady state,
the initial conditions and the initial perturbations are completely
lost in subsequent evolution. The seeds for the later shock instability
are the non-linear convective structures that arise in the first
post-bounce tens of milliseconds. Beyond these, we introduce no artificial
numerical perturbations.

\subsection{Microphysics}
\label{micro}

We employ the EOS of Shen et al. (1998), since it correctly
incorporates alpha particles and is more easily extended to lower densities
and higher entropies than the standard Lattimer \& Swesty (1991) EOS.
The neutrino-matter interaction physics is taken from Thompson, Burrows, \& Pinto (2003)
and Burrows \& Thompson (2004).  The tables generated in T/$\rho$/Y$_e$/neutrino-species
space incorporate all relevant scattering, absorption, and emission processes.
We follow separately the electron neutrino ($\nu_e$) and anti-electron neutrino ($\bar{\nu}_e$),
but for computational efficiency we lump the four remaining known neutrinos into ``$\nu_{\mu}$"
bins in the standard fashion.  Our baseline models have 16 energy groups for each species,
distributed logarithmically from 1 or 2.5 MeV to 250 or 320 MeV. As implied above,  
neutrino radiation pressure is handled consistently with a local ``$-\kappa F_{\nu} /c$"
body-force term in the momentum equation. Due to extreme matter-suppression 
effects, we have not felt it necessary to incorporate the effects of 
neutrino oscillations, but have developed a fully-consistent
formalism to do so  (Strack \& Burrows 2005).

\end{appendix}


\tabletypesize{\scriptsize}
\begin{center}
\begin{tabular}{cccc}
\multicolumn{4}{c}{Table 1. SASI Frequency versus Accretion Rate and Shock Radius$^{\dagger}$.}\\
\tableline\tableline
\bigskip
{Model (M$_{\odot}$)} & {  Frequency (Hz)} & {  $\dot{M}$ (M$_{\odot}$ s$^{-1}$)} & 
{  $<$R$_{shock}$$>$ (km)}\\
\medskip
WHW02-11.2 & 32 & 0.08 & 250 \\
WHW02-13 & 47 & 0.25 & 175 \\
WHW02-15 & 73 & 0.7 & 130 \\
WHW02-20 & 63 & 0.3 & 155 \\
WHW02-25 & 80 & 0.8 & 120 \\
\tableline
\end{tabular}
\end{center}
$^{\dagger}${$<$R$_{shock}$$>$ is the average shock radius after the SASI becomes
nonlinear, but before explosion.  The SASI frequency given is for the dominant 
shock oscillation component during this same time interval. $\dot{M}$ is near 
the average accretion rate onto the protoneutron star through a radius of 500 km
during this same phase.}

\clearpage

\end{document}